\documentclass[12pt,preprint]{aastex}
\usepackage{lscape}

\newcommand{\Mth}{M_{\rm th}}
\newcommand{\rs}{r_{\rm s}}
\newcommand{\phip}{\Phi_{\rm p}}
\newcommand{\cs}{c_{\rm s}}
\newcommand{\mplanet}{M_{\rm p}}
\newcommand{\ls}{l_{\rm sh}}

\def\ba{\begin{eqnarray}}
\def\ea{\end{eqnarray}}
\def\etal{et al.\ \rm}

\begin{document}
\title{Density Waves Excited by Low-Mass Planets in Protoplanetary Disks I: Linear Regime
}
\shorttitle{Linear Stage of the Density Waves in Protoplanetary Disc}

\shortauthors{Dong et al}

\author{Ruobing Dong, Roman R. Rafikov\altaffilmark{1}, James M. Stone, and Cristobal Petrovich}

\affil{\it Department of Astrophysical Sciences, Princeton University, Princeton, NJ 08544; rdong@astro.princeton.edu, rrr@astro.princeton.edu, jstone@astro.princeton.edu, cpetrovi@astro.princeton.edu}
\altaffiltext{1}{Sloan Fellow}

\begin{abstract}
Density waves excited by planets embedded in protoplanetary disks play a central role in planetary migration and gap opening processes. We carry out 2D shearing sheet simulations to study the linear regime of wave evolution with the grid-based code Athena, and provide detailed comparisons with the theoretical predictions. Low mass planets (down to $\sim 0.03M_\oplus$ at 1 AU) and high spatial resolution (256 grid points per scale height) are chosen to mitigate the effects of wave nonlinearity. To complement the existing numerical studies, we focus on the {\it primary} physical variables such as the spatial profile of the wave, torque density, and the angular momentum flux carried by the wave, instead of {\it secondary} quantities such as the planetary migration rate. Our results show {\it percent} level agreement with theory in both physical and Fourier space. New phenomena such as the change of the toque density sign far from the planet are discovered and discussed. Also, we explore the effect of the numerical algorithms, and find that a high order of accuracy, high resolution, and an accurate planetary potential are crucial to achieve good agreement with the theory. We find that the use of a too large time-step without properly resolving the dynamical time scale around the planet produces incorrect results, and may lead to spurious gap opening. Global simulations of planet migration and gap opening violating this requirement may be affected by spurious effects resulting in e.g. the incorrect planetary migration rate and gap opening mass. 
\end{abstract}



\section{INTRODUCTION}\label{sec:introduction}


The discovery of a number of ``hot Jupiters'' --- extrasolar giant planets residing on very tight orbits --- calls for understanding the dynamical pathways that brought these objects to their current orbits. Due to the difficulty of forming these planets {\it in situ}, it was suggested that potential hot Jupiter's form at larger separations from their stars and are subsequently transported inwards by some migration mechanism \citep{pap06,pap07}. In protoplanetary disks, planets act as sources of nonaxisymmetric density waves \citep[hereafter GT80]{gol79,gol80}, and coupling of planetary gravity to these waves acts to transfer angular momentum between the disk and the planet. This ultimately results in the orbital migration of embedded planets. At the same time, density waves generated by planets can affect the structure of the disk (potentially leading to gap opening) once the angular momentum that they carry is transferred to the disk material. Thus, better understanding of t
 he density wave properties and processes of their excitation, propagation, and damping plays a crucial role in the global picture of planet formation.

Evolution of the density wave can be separated into the linear and nonlinear stages. The former applies to density waves excited by low-mass planets, which have not had enough time to propagate very far from their launching site (a more precise condition is formulated in \S\ref{sec:theory}). The linear theory of disk-planet interaction was pioneered by \citet[GT80]{gol79} who provided a description of the density wave excitation in two-dimensional (2D) gaseous disks. This theory has been subsequently refined in various ways, e.g. by calculating the asymmetry of planetary interaction with inner and outer portions of the disk (Ward 1986), and extending it to fully three-dimensional disks \citep{tan02}.

These studies concentrated on understanding the behavior of perturbed quantities Fourier decomposed in azimuthal angle \citep{art93a,art93b,kor93}, often addressing only the {\it integral} wave properties such as the total angular momentum carried by the waves, or the differential Lindblad torque \citep{war86}. Only recently has the linear evolution of the wave properties in {\it physical space}, namely the spatial dependence of the perturbed fluid velocity and surface density, been addressed by \citet[hereafter GR01]{goo01} in the local shearing sheet approximation. Subsequently \citet{ogi02} and \citet{R02a} studied the shape of the density wake in the global setting. This paved way for subsequent exploration of the nonlinear wave evolution (GR01, \citealt{bate02}), understanding of which requires knowledge of the wave characteristics in physical space.

In parallel with these analytical and semi-analytical studies, numerous attempts to verify linear theory by direct hydrodynamical simulations were made. However, the majority of these studies compared with theoretical predictions only the {\it secondary} or derived properties of the planet-disk system related to the density wave excitation and evolution. For example, in their simulations \citet{lin93}, \citet{war97}, \citet{dan03}, and \citet{dan08} measure the rate of planet migration due to its gravitational interaction with the disk --- a quantity that is determined by the small difference between the summed (over all azimuthal harmonics) one-sided torques that the planet exerts on the inner and outer portions of the disk. Other studies \citep{lin93,mut10} tried to measure the planetary mass at which a gap opens in the disk --- a process that depends not only on the magnitude of the angular momentum carried by the waves but also on the intricate details of their dissipation. In both situations the verification of the linear theory ends up being rather indirect. 

It is certainly much more convincing and more robust to verify linear theory by comparing with its predictions the {\it primary} wave properties, such as the spatial distribution of the perturbed fluid properties derived from simulations. Recently some effort in this direction has been made. In particular, \citet{dan08} investigated the distribution of the torque imparted by the planet on the disk as a function of radial distance. \citet{mut10} showed the spatial structure of the density perturbation associated with the wave obtained in their 2D simulations. Unfortunately, in both cases no quantitative comparison with linear theory predictions was provided. 

In this work, we use 2D local inviscid hydrodynamical simulations in the shearing sheet geometry to verify the linear theory. For the first time we provide direct comparison of the primary physical variables --- perturbed surface density profile, torque density in both {\it physical} and {\it Fourier} space, and angular momentum flux --- derived from simulations with the theoretical predictions. Our verification effort is the most thorough, sophisticated and quantitative to date, and we achieve agreement with theory at the level of {\it several per cent} for all physical variables explored. In addition, using our numerical results we address some remaining issues in the linear theory, such as the different expressions for the torque density in Fourier space (GR80,\citealt{art93a}). We also formulate a set of requirements which should be satisfied by simulations to reach convergence and to reliably reproduce the linear theory results. The way in which we make the comparison an
 d the agreement we achieve can serve as a standard for checking the performance of different codes.

The paper is structured as follows. We briefly summarize the main results of the linear density wave theory in \S \ref{sec:theory}. A description of the code and our numerical setup are given in \S \ref{sec:code}, including a discussion of potential problems associated with implementing orbital advection algorithm in studies of the disk-satellite interaction. We present our numerical results and compare them with theoretical predictions in \S \ref{sec:result}, followed by the investigation of the sensitivity of simulation outcomes to the different numerical algorithms in \S \ref{sec:parameters}. In \S \ref{sec:fargo} we describe and discuss a commonly ignored numerical problem in disk simulations which if not handled appropriately can lead to incorrect results. Finally, a summary is provided in \S \ref{sec:summary}.


\section{Summary of linear theory results.} 
\label{sec:theory}


In linear theory of disk-planet interaction laid out in cylindrical geometry \citep[GT80]{gol79}, the planetary potential is normally decomposed into a series of Fourier harmonics. Assuming all perturbed fluid variables to be proportional to $\exp[i(m(\varphi-\Omega_pt)+\int^r k_r(r^\prime)d r^\prime)]$ ($m$ is a positive integer) and neglecting the self-gravity of the disk, one can obtain the dispersion relation for the $m$-th wave harmonic in the WKB limit in the form
\ba
m^2\left[\Omega(r)-\Omega_p\right]^2=\kappa^2+k_r^2c_s^2,
\label{eq:disp_rel}
\ea
where $\kappa$ is the epicyclic frequency ($\kappa=\Omega$ for a Newtonian potential), $k_r$ is the wavenumber, $\Omega$ is the fluid angular frequency, and $\Omega_p$ is the pattern speed of the perturbation equal to the angular frequency of the point-mass perturber moving on a circular orbit with semi-major axis $r_p$, that is $\Omega_p=\Omega(r_p)$. As Eq. (\ref{eq:disp_rel}) demonstrates, there are two (inner and outer) Lindblad resonances (LR) where $k_r\to 0$ and $m$-th harmonic of planetary potential couples best to the wave-like perturbation. These resonances are located at radii $r_{L,m}$ given by $\Omega(r_{L,m})=\Omega_pm/(m\pm 1)$, which in the limit of $m\gg 1$ results in
\begin{equation}
  x_{L,m}=\pm\frac{2h}{3\mu},
\label{eq:resonance-position}
\end{equation}
where $x\equiv r-r_p$, $h=c_s/\Omega_p$ is the vertical disk scale length, and $\mu\equiv mh/r_p$. 

The superposition of all the harmonics with different azimuthal wavenumbers $m$ forms inner and outer spiral density waves. In physical space fluid perturbation associated with the wave is concentrated in narrow wake (with azimuthal extent of order $h$), the location of which is described in the local limit by (GR01) 
\ba
y_w\approx -\mbox{sgn}(x)\frac{3}{4}\frac{x^2}{h}, 
\label{eq:wake_shape}
\ea
where $y\equiv r_p(\varphi-\varphi_p)$ and azimuthal angle $\varphi$ is counted in the prograde sense (with respect to planetary orbital motion). This location corresponds to the stationary phase of different Fourier harmonics composing the wave, as can be easily seen from Eq. (\ref{eq:disp_rel}). The wake shape in the global setting accounting for cylindrical geometry of the disk and Keplerian profile of $\Omega(r)$ was explored by \citet{ogi02} and \citet{R02a}. 

Each of the perturbation harmonics carries angular momentum flux, which for the $m$-th harmonic is given by (GT80).
\ba
&& F_H(m)=F_H^{\rm WKB}(m)\frac{F_H(m)}{F_H^{\rm WKB}(m)},
\label{eq:F_H}\\
&& F_H^{\rm WKB}(m)=\frac{4}{3}\frac{m^2\Sigma_0}{\Omega_p^2}
\left(\frac{G \mplanet}{r_p}\right)^2\left[2K_0(2/3)+K_1(2/3)\right]^2.
\label{eq:F_H_WKB}
\ea
Here $\Sigma_0$ is the unperturbed surface density of the disk,  
$F_H^{\rm WKB}(m)$ is the asymptotic Fourier contribution of the flux reached beyond the launching region on one side of the disk calculated in the WKB
approximation ($K_0$ and $K_1$ are the modified Bessel functions of
zeroth and first order), and the expression is only accurate for $m\gg1$. The ratio $F_H(m)/F_H^{\rm WKB}(m)$ accounts for 
the difference between the precise value of $F_H(m)$ and $F_H^{\rm WKB}(m)$. This difference 
becomes important for $m\gtrsim r_p/h\gg 1$ since, as demonstrated by
GT80, $F_H(m)$ decays exponentially with 
$m$ at high azimuthal wavenumbers --- the so-called ``torque cutoff'' 
phenomenon. As a result, the total angular momentum flux carried by 
the wave 
\ba
F_H=\sum_{m=1}^\infty F_H(m)
\approx 0.93\left(G \mplanet\right)^2\frac{\Sigma_0 r_p\Omega_p}{c_s^3},
\label{eq:F_H_tot}
\ea
is dominated by harmonics with $m\sim r_p/h$ (or $\mu\sim1$), which are excited within a radial separation of order $h$ from the planet, see Eq.
(\ref{eq:resonance-position}). This local nature of the wave excitation has important implications for the 
possibility of analytical treatment of the wave evolution.

In real protoplanetary disks, the torques acting on the inner and outer disks are slightly different (fractional difference $\sim h/r_p\ll 1$), giving rise to a net torque on the planet and leading to Type I planetary migration \citep{war86}. However, in shearing sheet configuration the inner and outer parts of the disk are identical so there is no net torque acting on the planet and migration does not arise. 

There is an important planetary mass scale characterizing excitation and propagation of the density waves, the so called {\it thermal mass} (GR01), which we define\footnote{This definition differs from that of GR01 who used $M_1=2\cs^3/(3\Omega_p G)=(2/3)\Mth$ instead of $\Mth$.} here as 
\ba
\Mth=\frac{c_s^3}{G\Omega_{\rm p}}\approx 12~M_\oplus
\left(\frac{c_s}{1~\mbox{km s}^{-1}}\right)^{3}
\left(\frac{M_\odot}{M_\star}\right)^{1/2}
\left(\frac{r_p}{\rm AU}\right)^{3/2}.
\label{eq:Mth}
\ea
There are several reasons for the significance of this mass scale. First, a Hill radius of the planet with $\mplanet=\Mth$ is of order the scale height of the gaseous disk $h$. Second, the Bondi radius $R_B\equiv G\mplanet/c_s^2$, which is the size of a region in which planetary gravity strongly perturbs local pressure distribution, also becomes comparable to $h$ for $\mplanet=\Mth$ (Rafikov 2006). Third, and most importantly for our present study, planets with $\mplanet\gtrsim \Mth$ generate density waves which are nonlinear from the very start ($\delta \Sigma/\Sigma_0\gtrsim 1$). 

In the small planet mass limit of 
\ba
\mplanet\lesssim\Mth
\label{eq:limit}
\ea
perturbations induced by the planetary gravity are weak ($\delta \Sigma/\Sigma\lesssim 1$) and {\it excitation} of the density wave by the planet can be studied purely in the linear approximation. It is in this limit that the linear theory of wave {\it driving} developed by GT80 and all of its results are valid. In the opposite limit of $\mplanet\gtrsim\Mth$ the nonlinear effects must be accounted for even at the stage of wave excitation, which is not feasible analytically. 

As mentioned above, the generation of the wave is largely completed at a distance of order $h$ away from the planet, because this is where the most important Lindblad resonances lie. After that, in the absence of viscosity, the wave travels freely, no longer affected by the planetary gravity. At this stage of {\it propagation} nonlinear effects start to accumulate and finally result in wave breaking, formation of a shock, and transfer of wave energy and angular momentum to the disk fluid. Prior to the steepening into a shock the angular momentum flux $F_H$ carried by the wave far from the planet (outside of the excitation region $|x|\sim h$) is exactly conserved in the absence of linear dissipation. This nonlinear stage has been explored analytically in GR01 in the limit (\ref{eq:limit}). The numerical results on the nonlinear wave evolution will be presented in \citet[here after Paper II]{don11}.

We do not cover this stage in this work, but we mention the result of GR01 for the radial separation away from the planetary orbit at which the shock induced by the nonlinear effects first occurs:
\begin{equation}
\ls\approx 0.8 
\left(\frac{\gamma+1}{12/5}\frac{ \mplanet}{\Mth}\right)^{-2/5}h,
\label{eq:ls}
\end{equation}
where $\gamma$ is the adiabatic index of gas. For $\mplanet\ll \Mth$ one finds the shock to be well separated from the planet, $\ls\gtrsim h$, so that the wave excitation occurs in the linear regime, as asserted before. However, even for $\mplanet\ll \Mth$ the linear theory of wave propagation fails for $x\gtrsim \ls$. In this work we are always exploring the low planetary mass limit (\ref{eq:limit}) and consider only $|x|\lesssim \ls$ so that linear theory applies.


\section{Numerical setup}
\label{sec:code}


Most analytical work on the linear evolution of density waves was carried out in application to two-dimensional (2D) disks. To provide a meaningful comparison with these studies, in our current work we employ 2D hydrodynamical simulations in local shearing box configuration. The computational tool for this investigation is Athena, a grid-based code for astrophysical gas dynamics using higher-order Godunov methods. Athena is written in conservative form, so it conserves mass, momentum, and energy down to machine precision. The mathematical foundations of the algorithms are described in \citet{gar05,gar08}, and a comprehensive description of the implementation and tests of the algorithms is given in \citet{sto08}. 

The implementation of the shearing box approximation in Athena is described in \citet{sto10}. This approximation adopts a frame of reference located at radius $r_p$ corotating with the disk at orbital frequency $\Omega_p=\Omega(r_p)$.  In this frame, the 2D hydrodynamics equations are written in a Cartesian coordinate system $(x,y,z)$ that has unit vectors ${\bf \hat{i}}$, ${\bf \hat{j}}$, and ${\bf \hat{k}}$ as \citep{sto10}:
\ba
&& \frac{\partial \Sigma}{\partial t} + {\bf\nabla\cdot} (\Sigma{\bf v}) = 0,
\label{eq:cons_mass} \\
&& \frac{\partial \Sigma {\bf v}}{\partial t} + {\bf\nabla\cdot} (\Sigma{\bf vv} + p) = \Sigma \Omega_p^{2}(2qx{\bf \hat{i}}) - 2\Omega_p {\bf \hat{k}} \times \Sigma {\bf v} - \Sigma\nabla\Phi_p,
\label{eq:cons_momentum} \\
&& \frac{\partial E}{\partial t} + \nabla\cdot (E{\bf v} + p \cdot {\bf v}) = \Omega_p^{2}\Sigma{\bf v} \cdot (2qx{\bf \hat{i}}) - \Sigma{\bf v}\cdot\nabla\Phi_p,
\label{eq:cons_energy}
\ea
where $\phip$ is the gravitational potential of the planet (whose form we will discuss in the next section), $p$ is the gas pressure, $E$ is the total energy density (sum of the internal and kinetic energy, but excluding the gravitational energy). The shear parameter $q$ is defined as:
\begin{equation}
 q = - \frac{d {\rm ln} \Omega}{d {\rm ln} r}
\label{eq:shearparam}
\end{equation}
so that for Keplerian flow $q=3/2$. The equation of state (EOS) we use in the simulations is the ideal gas law with
\begin{equation}
  E_{\rm in} = \frac{p}{\gamma -1}
\label{eq:ein}
\end{equation}
where $E_{\rm in}$ is the internal energy and $\gamma=5/3$. We also use an isothermal EOS, in which case $p=\Sigma\cs^2$, where $\cs$ is the isothermal sound speed. We do not include explicit viscosity (i.e. there is no explicit linear dissipation in the system) and we do not account for the self-gravity of the disk (thus we concentrate on studying the low mass disks). 

To accurately resolve the flow in the vicinity of a gravitating mass without making the time step too short one usually resorts to softening the potential of the perturber. There are many ways of doing this but all of them come up with the form of the potential that converges to the Newtonian potential $\Phi_K=-G\mplanet/\rho$ at large separations $\rho=\sqrt{x^2+y^2}$, beyond the softening length $\rs$. At small separations, $\rho\ll \rs$ smoothed potentials become finite allowing the time step of simulations to stay finite. To test the sensitivity of our results to the specific method of potential softening we tried three different forms of the softened planetary potential. The second order potential 
\ba
\phip^{(2)}=-G \mplanet\frac{1}{(\rho^2+\rs^2)^{1/2}}
\label{eq:phi2}
\ea
converges to $\Phi_K$ at $\rho\gg \rs$ as $\left(\rs/\rho\right)^2$ (which means the fractional error is O$((r_s/\rho)^2)$ as $r_s/\rho\rightarrow0$). This is a standard form of the potential used in a majority of numerical hydrodynamic studies. We also studied the fourth order potential
\ba
\phip^{(4)}=-G \mplanet\frac{\rho^2+3\rs^2/2}{(\rho^2+\rs^2)^{3/2}}
\label{eq:phi4}
\ea
converging to the point mass potential as $\left(\rs/\rho\right)^4$ for $\rho\gg \rs$, and the sixth order potential 
\ba
\phip^{(6)}=-G \mplanet\frac{\rho^4+5\rho^2\rs^2/2+15\rs^4/8}{(\rho^2+\rs^2)^{5/2}}
\label{eq:phi6}
\ea 
converging to $\Phi_K$ at $\rho\gg \rs$ as $\left(\rs/\rho\right)^6$. The difference in accuracy with which these potentials represent $\Phi_K$ at large $\rho$ can be quantified by the distance from the planet at which a given potential deviates from $\Phi_K$ by $1\%$. This distance is $7\rs$ for $\phip^{(2)}$, $2.3\rs$ for $\phip^{(4)}$, and $1.5\rs$ for $\phip^{(6)}$. Note that the first-order potential $\phip^{(1)}=-G \mplanet/|\rho+\rs|$, which we do not use here, gives only $10\%$ accuracy relative to $\Phi_K$ even at $\rho=10\rs$, and we strongly discourage its use. The effects of different potential prescriptions will be discussed in detail in \S \ref{sec:phip+rs}.

To guarantee accuracy and convergence of our results we carry out an exhaustive exploration of various numerical parameters characterizing our simulations, described in \S \ref{sec:parameters}. Unless noted otherwise, our results shown in \S \ref{sec:result} are obtained with the following numerical parameters: an isothermal equation of state, the Roe solver with third order reconstruction in characteristic variables, and the corner transport upwind (CTU) unsplit integrator \citep{sto08}, a resolution of 256 grid points per scale length $h$ (subsequently denoted $256/h$ for brevity), and fourth order accurate potential $\phip^{(4)}$ given by Eq. (\ref{eq:phi4}) with softening length $\rs=h/32$.

High order of accuracy, high resolution and an accurate form of planetary potential used in our runs allow us to properly capture the details of wave evolution. High resolution ensures low levels of numerical viscosity and prevents the angular momentum accumulated by the wave from spurious dissipation (see Section \ref{sec:parameters} for more discussion). We run a series of test simulations with otherwise identical conditions but different explicit Navier-Stokes viscosity, and measure the wave properties. As the explicit viscosity decreases the simulation results gradually converge to the one with zero explicit viscosity, which indicates the numerical viscosity dominates the explicit viscosity. For a typical simulation with low $\mplanet=2.09\times 10^{-2}\Mth$ and an isothermal equation of state, the effective Shakura-Sunyaev $\alpha$-parameter ($\alpha=\nu/h\cs$) characterizing our numerical viscosity is found to be below $10^{-5}$. Such small levels of viscosity are expected in dead zones of protoplanetary disks \citep{gam96}, where magnetorotational instability (MRI) may not operate effectively \citep{flem03}. Also note even when the MRI operates, MHD turbulence may not act like a Navier-Stokes viscosity.

\begin{figure}[tb]
\begin{center}
\epsscale{0.75} \plotone{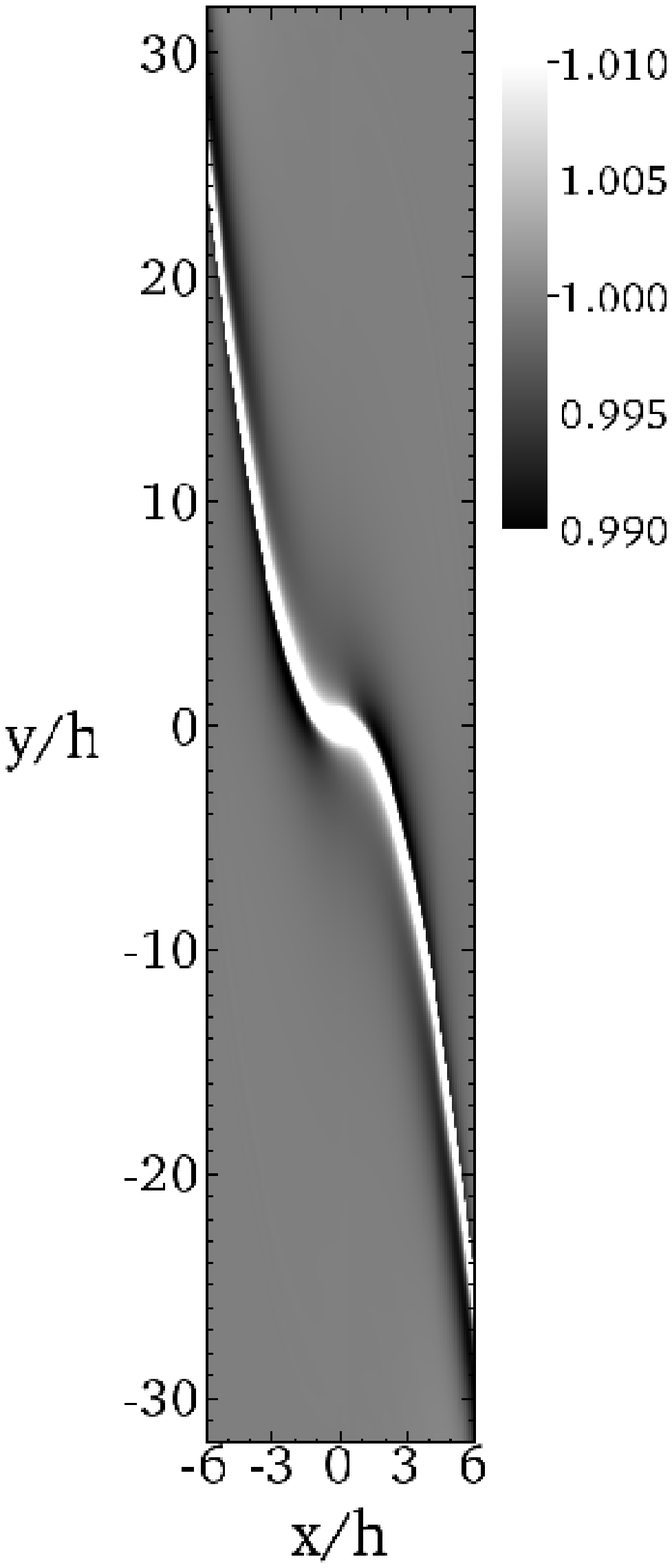}
\end{center}
\figcaption{A typical image of our simulations, showing the density structure and the spiral waves. The quantity plotted here is $\Sigma/\Sigma_0$. $\mplanet=2.09\times10^{-2}\Mth$ in this case.
\label{fig:image}}
\end{figure}

For the standard simulations in this work to fit the parabolic wake (see Eq. \ref{eq:wake_shape}) in the box we use box size $12h\times64h$, thus the overall grid resolution in our runs is about $3072\times 16384$ (for the standard resolution of $256/h$). In a few cases with very small $\mplanet$, we extend the simulation box size to $20h\times156h$ to trace linear wave evolution further out in $x$, since smaller $\mplanet$ delays shock formation (see eq. (\ref{eq:ls})). Our simulations are run for at least 10 and in some cases up to 50 orbital periods. Figure \ref{fig:image} presents a typical snapshot of one of our simulations showing the density structure and the spiral waves. 

We use the following boundary conditions (BCs). On $x$ (radial) boundaries, we keep values of all physical variables in ghost zones fixed at their respective unperturbed Keplerian values ({\it i.e.} keep the ghost zones as their initial states), and the waves leave through the $x$ boundaries when they reach the edge. In our shearing sheet simulations, experiments show that this $x$ BC has less wave reflection and outflowing of the fluid than the conventional outflow $x$ BC, in which case the ghost zones are copied from the last actively-updated column of cells of the simulations at every time-step (see also \citealt{mut10}). We do not expect our adopted radial BC to affect wave evolution since significant radial fluid motions are not expected to arise in our simulations anyway.

On the $y$ (azimuthal) boundary, we experimented with two BCs: the conventional outflow BC, as described above, and an inflow/outflow BC. In the latter case, the variables in the ghost zones are fixed at their initial values if they are the physical ``inflow'' boundaries (the regimes $y<0, x<0$ and $y>0, x>0$), or copied from the last actively-updated row of cells if they are the physical ``outflow'' boundaries (the regimes $y>0, x<0$ and $y<0, x>0$). For the purposes of current paper, the difference between these two BCs is not significant and resultant density profiles and torque calculation are almost identical (of the level of $10^{-3}$). We found that with the pure outflow BC fluid entering the simulation box accumulates some non-zero velocity perturbation on top of the pure linear shear velocity profile. This affects calculation of variables derived from the simulated velocity field, such as potential vorticity, which are analyzed in Paper II. We use the inflow/outflow 
 $y$ BC for our simulations.

At the start of a simulation run we instantaneously turn on the potential of the planet in the center of the linearly sheared fluid flow with uniform surface density. This gravitational perturbation immediately excites an inner and an outer density wave propagating away from the planet, as well as strong transients: vortices that appear near the planet but travel on horseshoe orbits away from it. Before measuring the wave properties, we run simulations for $\sim15$ orbital periods to let these time-dependent structures move away from the planet. We also tried gradually turning on planetary gravity by linearly increasing strength of the potential from zero to its full value within several orbital periods \citep{mut10,li09}, but did not find this trick as effective at removing transient structures as simply waiting for them to move away from the planet.


\section{Result and comparison with linear theory}
\label{sec:result}


We now present our main results and compare them with the predictions of linear theory. To mitigate possible nonlinear effects we consider only very low planetary masses in our calculations, ranging between $3.2\times 10^{-2}\Mth$ and $2.8\times 10^{-3}\Mth$. According to Eq. (\ref{eq:Mth}) at 1 AU this mass range corresponds to $\mplanet$ between $0.4$ M$_\oplus$ and 3 Lunar masses (note that they are sound speed dependent). As the basis for comparison we primarily use the distribution of the perturbed density (\S\ref{sec:density}) and the evolution of the angular momentum flux carried by the density wave (\S\ref{sec:amf}). We carry out linear theory verification both in physical (coordinate) space and in Fourier space.  


\subsection{The density wave profile}
\label{sec:density}

The most basic and direct comparison between numerical simulations and theory can be performed using the spatial distribution of the perturbed fluid variables. In this work we choose the perturbed surface density $\delta\Sigma\equiv \Sigma-\Sigma_0$ as the primary variable for comparison. Since $\delta\Sigma$ is spatially concentrated along a narrow wake it makes sense to compare with theory both the {\it overall shape} of the wake in $x-y$ coordinates, and the density distribution {\it across} the wake.

\begin{figure}[tb]
\begin{center}
\epsscale{0.45} \plotone{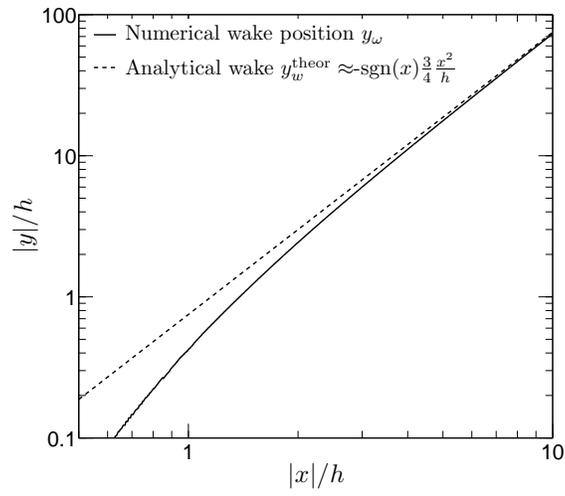}
\end{center}
\figcaption{Location of the peak density in the wake in the $x-y$ plane measured from simulations ({\it solid curve}), compared to the analytical wake shape ({\it dashed curve}, Eq. (\ref{eq:wake_shape})). As expected, they agree far from the planet. Numerical peak density position is nearly independent of the simulation parameters. 
\label{fig:wake}}
\end{figure}

We determine the wake shape in the following way. At each value of $x$ we find the value of $y$, called $y_w(x)$, at which $\delta\Sigma$ reaches its maximum value $\delta\Sigma_{max}(x)$. We then compare the run of $y_w(x)$ with the theoretical prediction (\ref{eq:wake_shape}), which applies in the shearing sheet geometry (GR01). The results are shown in Figure~\ref{fig:wake}. The two curves agree well far from the planet, at $|x|\gtrsim h$. The discrepancy at $|x|\lesssim h$ is not surprising because in this region (a) the density wave is not yet fully formed and (b) the dispersion relation (\ref{eq:disp_rel}) is significantly affected by the $\kappa^2$ term, which causes the wake profile to deviate from the theoretical prediction (\ref{eq:wake_shape}) valid far from the planet, see \citet{ogi02} for details.

\begin{figure}[tb]
\begin{center}
\epsscale{0.45} \plotone{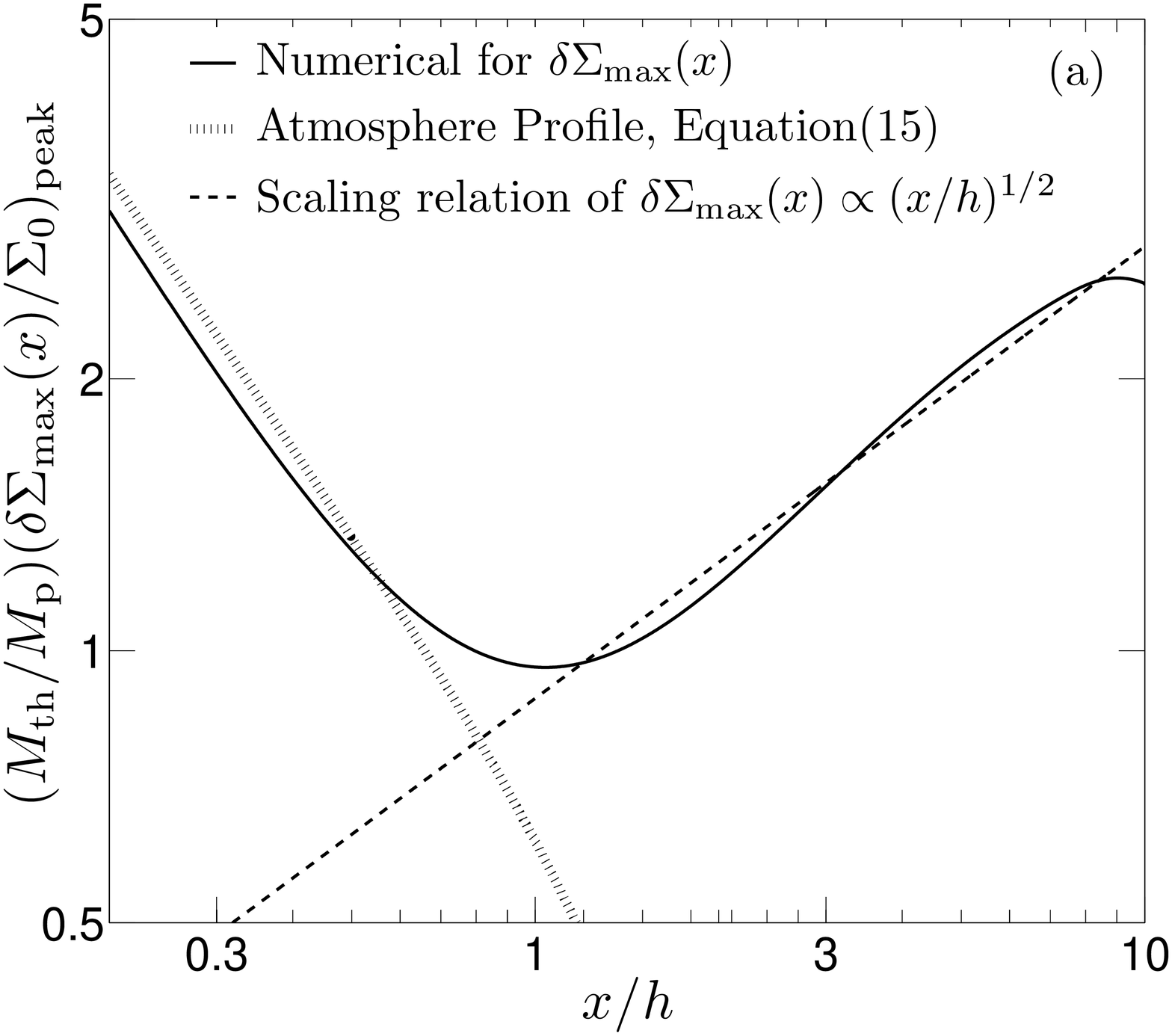} \plotone{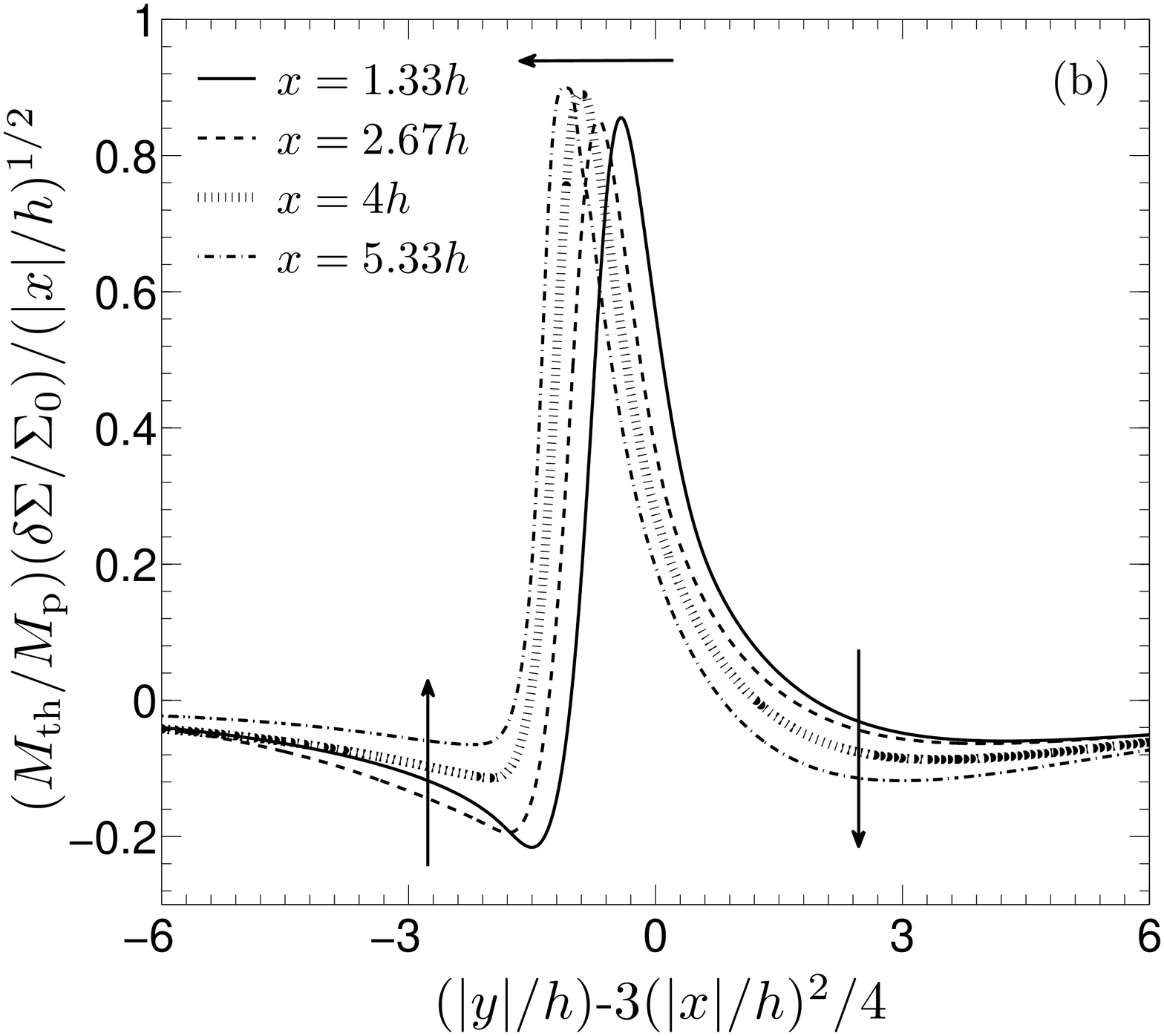}
\end{center}
\figcaption{Behavior of the density perturbation in a simulation with the following parameters: $\mplanet=3.7\times 10^{-3}\Mth$ (corresponding to $\approx 3.6$ Lunar masses at 1 AU), shocking distance $\ls\approx 8h$. (a) Variation of the peak value of the relative density perturbation $\delta\Sigma/\Sigma_0$ with $x$ ({\it solid curve}). Analytical profile (\ref{eq:atmosphere}) of the quasi-static gaseous envelope collected inside the planetary potential well (without background shear) is shown by the dotted curve. Dashed line shows theoretical scaling $\delta\Sigma\propto x^{1/2}$ (normalization is arbitrary) resulting from conservation of the angular momentum flux carried by the wave. (b) Azimuthal density profiles $\delta\Sigma$ (scaled by the planetary mass and normalized by $(x/h)^{1/2}$) at several values of $x$. To be compared with theoretical density profiles computed in linear approximation at the same locations in GR01 (their Figure 1). See text for details.
\label{fig:density-profile}}
\end{figure}

Next we investigate the evolution of the wave amplitude as it travels away from the planet by looking at the behavior of maximum amplitude $\delta\Sigma_{max}(x)\equiv \delta\Sigma(x,y_w(x))$ as a function of $x$. We plot this quantity scaled by $\Sigma_0(\mplanet/\Mth)$ (this normalization removes the dependence of $\delta\Sigma_{max}$ on $\Sigma_0$ and $\mplanet$) in Figure \ref{fig:density-profile}a. 

At small $x\lesssim h$ the value of $\delta\Sigma_{max}$ is large and decreases with $x$. This behavior has nothing to do with the density wave since the region in the immediate vicinity of the planet represents a quasi-static atmosphere that forms inside the planetary potential well right after the planetary gravity is switched on. This structure is rarely mentioned in the analytical calculations of planet-disk interaction but it shows prominently in realistic simulations. In hydrostatic balance (in the absence of background velocity caused by differential rotation) density profile in the vicinity of the planet should be given by (assuming isothermal equation of state with constant sound speed $c_s$)
\begin{equation}
\Sigma_{\rm atm}(\rho)=\Sigma_0 e^{-\phip(\rho)/c_s^2}
\label{eq:atmosphere}
\end{equation}
In real disks, the background velocity field distorts the atmospheric profile from the circularly-symmetric (with respect to planet) shape predicted by this equation. This explains why $\Sigma_{\rm atm}(x)$ shown Figure \ref{fig:density-profile}b slightly overestimates $\delta\Sigma_{max}$ at small $x$. 

The effect of the atmosphere on $\delta\Sigma_{max}$ rapidly decreases beyond several $h$ from the planet, and $\delta\Sigma_{max}$ starts to rise with increasing $x$. This is because density wave excitation occurs at $|x|\sim h$ and from now on the behavior of $\delta\Sigma_{max}$ is determined by the wave-like density perturbation. Conservation of the angular momentum flux $F_{\rm H}$ carried by the wave forces the wave amplitude to increase with distance $\propto x^{1/2}$ (GR01). As Figure \ref{fig:density-profile}a shows this scaling agrees reasonably well with the numerically computed $\delta\Sigma_{max}(x)$ far from the planet. 

Note that initially, at separations of several $h$ from the planet, $\delta\Sigma_{max}$ increases with $x$ {\it faster} than the theoretical $x^{1/2}$ scaling, which is caused by the still ongoing accumulation of the angular momentum flux $F_H$ at these separations. In other words, at this location $F_H$ has not yet reached its asymptotic value given by Eq. (\ref{eq:F_H_tot}), see \S\ref{sec:amf} for more details. 

Far from the planet, beyond $x=\ls$ ($\ls=7.5h$ in the case shown in Figure \ref{fig:density-profile}a), the peak $\delta\Sigma$ rapidly goes down relative to the analytical $x^{1/2}$ scaling. This behavior is caused by the appearance of the shock at $\ls$ and subsequent dissipation of the angular momentum flux carried by the wave. These nonlinear processes are explored in detail in Paper II.

Finally, we examine evolution of the density profile across the wake as a function of radial separation $x$. We do this by making an azimuthal cut through the density field at fixed $x$ and shifting the resultant one-dimensional density profile in $y$ by $y_w$ given in Eq. (\ref{eq:wake_shape}). To eliminate the increase of $\delta\Sigma$ resulting simply from the angular momentum flux conservation we normalize the density perturbation by $x^{1/2}$. We additionally scale $\delta\Sigma$ by $\Sigma_0(\mplanet/\Mth)$, as in Figure \ref{fig:density-profile}a. As mentioned before, in linear theory the normalized density profile should be independent of $\mplanet$ at a fixed separation $x$. This is indeed seen in Figure \ref{fig:mp}a, where we plot scaled profiles of $\delta\Sigma$ computed for several values of $\mplanet$ at $x=1.33h$, and find no significant difference between them. The adopted value of $x$ is small enough for the linear theory to apply even for the highest $\mplanet$ ($\approx 0.03\Mth$) used in making this Figure, i.e. $x$ is always considerably smaller than $\ls$ (cf. Figure \ref{fig:density-profile}b and the discussion in \S\ref{sec:nonlinear}). 

Previously, GR01 computed the evolution of the density profile as a function of $x$ in the linear regime. In Figure \ref{fig:density-profile}b we provide an analogue of Figure 1 of GR01 by showing the scaled density cuts at the same values\footnote{Note that GR01 normalized $x$ by the ``Mach 1''distance $l=(2/3)h$ rather than $h$ as we do here. However, the physical values of $x$ for the density cuts shown in Figure \ref{fig:density-profile}a are the same as in GR01.} of $x$ --- $1.33h$, $2.67h$, $4h$, $5.33h$ --- as in GR01. This particular calculation uses a very small planet mass $\mplanet=3.7\times 10^{-3}\Mth$ corresponding to $l_s\approx 8h$, see Eq. (\ref{eq:ls}), which puts the values of $x$ used in making this Figure well inside the shock position and reduces the impact of nonlinearity on the density profile, see \S\ref{sec:nonlinear}. 

The effective width of the density profile in this Figure does not significantly vary with $x$ and stays at the level of several $h$ at all times. We note, however, that this property holds only for the density cuts passing through the wake at fixed $x$, as shown in Figure \ref{fig:density-profile}b. On the contrary, the width of the density profile cutting through the wake at fixed $y$ (not shown here) is {\it shrinking} as $|x|^{-1}$, which follows directly from the dispersion relation (\ref{eq:disp_rel}) since $k_r\propto |x|$ for $x\gtrsim h$. 

Our numerical results agree with linear calculations of GR01 quite well, with quantitative differences at the level of $10\%$ or less. In particular, we are able to reproduce several subtle features of the wake profile evolution found in GR01, which are highlighted by arrows in Figure \ref{fig:density-profile}b. These are (a) the increase of $\delta\Sigma$ with $x$ in the density trough in front of the wake, (b) the decrease of $\delta\Sigma$ with $x$ in the density trough behind the wake, and (c) the slight drift of density profile with respect to theoretical wake position $y=y_w(x)$ given by Eq. (\ref{eq:wake_shape}) as $x$ increases. There are also some minor differences between the Figure \ref{fig:density-profile}b and the results presented in GR01, e.g. the lower (by about $10\%$) amplitude of $\delta\Sigma$ far from the planet, at $x\gtrsim3h$, in our case (note that the vertical axis of \ref{fig:density-profile}b is different from the vertical axis in Figure 2 of GR01, since we use $\Mth$ and $h$ instead of $M_1$ and $l=2h/3$ as mass and length units). We speculate that this may be explained by the gradually accumulating nonlinear effects in our simulations, which are absent in linear calculations of GR01.

Previously, \citet[Figure 6]{mut10} have shown density profiles resulting from their 2D hydro simulations in cylindrical geometry, which are similar to those presented in Figure \ref{fig:density-profile}. However, their calculations were carried out for planetary masses ({\it minimum} $\mplanet=5\times 10^{-2}\Mth$) larger than used in our work ({\it maximum} $\mplanet=3.2\times 10^{-2}\Mth$), which does not allow them to isolate linear effects clearly. Also, \citet{mut10} did not attempt quantitative comparison of their numerical results with the linear theory predictions.


\subsection{The angular momentum flux and torque density}
\label{sec:amf}

We now investigate the behavior of the angular momentum flux (AMF) carried by the wave $F_H(x)$ as a function of $x$. Linear calculations of this quantity in Fourier space have been first performed in GT80, and later refined  in \citet{art93a,art93b} and \citet{kor93}. The torque density $dT_H/dx$, which is a spatial derivative of $F_H(x)$ has been inferred from simulations in \citet{bate03} and \citet{dan08,dan10}, although no direct comparison with the linear theory was provided in these studies.

The analytical expression (\ref{eq:F_H_tot}) for the integrated angular momentum flux scales linearly with the planetary semi-major axis $r_p$. In the shearing sheet geometry employed in our simulations $r_p$ is ill-defined, and it makes sense to redefine angular momentum flux as $\tilde F_H\equiv F_H/r_p$. Subsequently we will drop tilde for brevity and denote such normalized momentum flux as $F_H$. We compute $F_H(x)$ from our data according to the following definition:
\begin{equation}
F_H(x) = \Sigma_0\int\limits_{-\infty}^\infty
dy\delta v_y v_x,
\label{eq:amf-numerical}
\end{equation}
where $v_x$ is the radial velocity of the fluid, and $\delta v_y$ is the velocity perturbation with respect to the background shear profile in the azimuthal direction.

We compute the torque density (normalized by $r_p$) exerted by the planet on the fluid at separation $x$ as
\begin{equation}
\frac{dT_H}{dx} =-\int\limits_{-\infty}^\infty dy\delta\Sigma\frac{\partial\phip}{\partial y},
\label{eq:torque-density}
\end{equation}
and the integrated torque accumulated by the wave at $x$ is just
\begin{equation}
T_H(x) = \int\limits_0^x\frac{dT_H}{dx}dx.
\label{eq:specific torque}
\end{equation}
Conservation of angular momentum ensures that the derivative of Equation (\ref{eq:amf-numerical}) (the slope of the AMF) coincides with the torque density (\ref{eq:specific torque}) in the absence of dissipation.


\subsubsection{Comparison with theory in Fourier space}
\label{sec:Fourier}

We first compare our simulation results with linear theory in Fourier space by studying the behavior of the torque cutoff function $F_H(m)/F_H^{WKB}(m)$, see \S\ref{sec:theory}. Previously GT80 computed variation of this quantity as a function of $m$, and the result is shown in Figures 2 and 3 of their paper, providing a basis for comparison. \citet{art93a} subsequently refined this calculation by arguing that evaluating the planetary potential at separations slightly displaced from the classical Lindblad resonance position (\ref{eq:resonance-position}) should lead to more accurate results. Adopting this approach \citet{art93b} provided a simple analytical prescription (Eq. (25) of \citealt{art93b}) to reproduce the behavior of $F_H(m)/F_H^{WKB}(m)$, which can also be compared with our calculations.

Substituting $v_x$ and $\delta v_y$ in the form of their Fourier integrals in Eq. (\ref{eq:amf-numerical}) and manipulating the resulting expression one can write $F_H(x)$ as the integral over the azimuthal wavenumber $k_y$:
\ba
&& F_H(x)=\int\limits_0^{\infty}dk_y F_{H,k}(x),\\
&& F_{H,k}(x)=4\pi\Sigma_0\left[{\rm Re}\left(v_{x,k}\right){\rm Re}\left(\delta v_{y,k}\right)+{\rm Im}\left(v_{x,k}\right){\rm Im}\left(\delta v_{y,k}\right)\right],
\label{eq:flux_Fourier}
\ea
where $v_{x,k}$ and $\delta v_{y,k}$ are the Fourier transforms of $v_x$ and $\delta v_y$, which themselves are functions of $x$. Even though the available analytical calculations of the torque cutoff function were performed only in the limit $x\gg 1$, we chose to retain the dependence of $F_H(x)$ on $x$ to explore the evolution of the harmonic content of the AMF with distance from the planet.

Instead of the discrete theoretical WKB Fourier harmonics of the flux $F_H^{WKB}(m)$ given by (\ref{eq:F_H_WKB}) we use the following continuous version\footnote{Transition between the discrete and integral representations of the AMF is performed by replacing $m$ with $k_yr_p$ and summation over $m$ with integration over $r_pdk_y$.}: 
\ba
F_{H,k}^{WKB}=\frac{4}{3}\Sigma_0\left(k_yh\right)^2
\left(\frac{G \mplanet}{c_s}\right)^2\left[2K_0(2/3)+K_1(2/3)\right]^2,
\label{eq:flux_Fourier_theor}
\ea
where we also normalized the final expression by $r_p$. 

We can now compute real and imaginary components of $v_{x,k}$ and $\delta v_{y,k}$ using our numerical data and then obtain $F_{H,k}(x)$ from Eq. (\ref{eq:flux_Fourier}). Dividing the result by $F_{H,k}^{WKB}$ provides us with the ratio $F_{H,k}(x)/F_{H,k}^{WKB}$, which can then be compared with the existing theoretical calculations of the torque cutoff. This is done in Figure \ref{fig:amf-decay} where we plot the numerical $F_{H,k}(x)/F_{H,k}^{WKB}$ computed at different separations from the planet against the semi-analytical calculations of the same quantity (in the limit $|x|\gg h$) performed by GT80 and \citet{art93b}. These calculations were done for rather small $\mplanet=1.2\times 10^{-2}\Mth$ (corresponding to $\ls\approx 5h$) allowing us to see linear wave evolution out to large separations.

One can see in Figure \ref{fig:amf-decay}a that the agreement between theory and simulations is generally quite good. Our results seem to agree better with GT80 torque cutoff prescription than with \citet{art93b}. However, at the level of accuracy available to us we are not able to firmly discriminate between these two torque cutoff prescriptions. At intermediate values of $\mu=k_yh \approx 0.5-5$ our numerical results for $F_{H,k}(x)/F_{H,k}^{WKB}$ are essentially independent of $x$ and pass between the two aforementioned analytical torque cutoff prescriptions. Only weak sensitivity of $F_{H,k}(x)/F_{H,k}^{WKB}$ on $x$ is expected since beyond $|x|=2h$ the power in azimuthal harmonics corresponding to this range of $k_yh$ should have already been fully accumulated by the density wave, and no evolution should arise. Good agreement with the GT80 results is also illustrated in Figure \ref{fig:amf-decay}b, which shows the torque cutoff function (multiplied by $\mu^2$) in linear 
 space, as in Figure 3 of GT80.

The situation is different for smaller values of $k_yh\lesssim 0.5$, where the numerical curves fall below the analytical asymptotic behavior $F_{H,k}/F_{H,k}^{WKB}\to 1$ as $k_yh\to 0$. Also, there is noticeable evolution of $F_{H,k}(x)/F_{H,k}^{WKB}$ with $x$, with nonzero power extending down to smaller and smaller values of $k_y$ as the wave travels further from the planet. This is because the low-order harmonics corresponding to small $k_yh$ contribute to the wave flux predominantly at separations larger than $\sim h$ \citep{kor93}, see Eq. (\ref{eq:resonance-position}). As a result, the further the wave travels, the more power gets collected by the wave from these lower-order azimuthal harmonics of the planetary potential, but the very small values of $k_y\lesssim 0.05$ never contribute to the wave flux even at $|x|=5h$. 

Our numerical results also exhibit noticeable disagreement with theory and clear evolution with $x$ at large values of $k_yh\gtrsim 5$, and we comment on their origin in \S\ref{sec:nonlinear}.


\subsubsection{Comparison with theory in physical space}
\label{sec:physical}

We now study the behavior of the AMF and torque density in physical space. Understanding the spatial distribution of the latter quantity is important for properly computing the AMF in disks with non-uniform distribution of $\Sigma$ in the vicinity of the planet, and is thus important for understanding the early stages of gap formation by massive planets. 

Here we look at disks with uniform distribution of $\Sigma$. In Figure \ref{fig:amf} we show the behavior of the AFM $F_H(x)$ computed according to the definition (\ref{eq:amf-numerical}) and also of the accumulated torque $T_H(x)$ given by Eqs. (\ref{eq:specific torque}). We display our results for four different values of $\mplanet$, normalizing $F_H(x)$ and $T_H(x)$ by $\Sigma_0(G\mplanet)^2/(h\cs^2)$ --- according to linear theory the shape of the resultant curves should then be independent of $\mplanet$. Theoretical scaling of the torque $\propto\mplanet^2$ is indeed largely confirmed by this Figure, but see \S\ref{sec:nonlinear} for more details.

To facilitate the comparison with the torque calculation, AMF curves have all been vertically shifted by a small offset to cancel the non-zero value of $F_H(0)$ at $x=0$ resulting from the residual gas motion in the horseshoe region and a small intrinsic non-zero starting value of the AMF (Rafikov \& Petrovich, in preparation). This brings them in perfect accord with $T_H(x)$ curves close to the planet, which is expected to be the case in the absence of dissipation\footnote{This additionally confirms low levels of numerical viscosity in our runs.}. At large separations, beyond $\ls$, the $F_H(x)$ curve starts falling below the $T_H(x)$ curve, which is caused by shock formation and dissipation of the wave AMF past this point. Quite naturally, this effect is more pronounced for higher $\mplanet$, e.g. $\mplanet=3.2\times10^{-3}\Mth$ and $\mplanet=1.2\times10^{-3}\Mth$, corresponding to smaller $\ls$. For smaller masses shocks occur outside the simulation box and essentially no difference between $F_H(x)$ and $T_H(x)$ can be seen. The post-shock behavior of the AMF will be studied in Paper II.

In Figure \ref{fig:torque-density} we plot the torque density $dT_H(x)/dx$ (\ref{eq:torque-density}) as a function of $x$. The overall shape of the curve is consistent with previous numerical calculations of the same quantity in cylindrical coordinates performed by \citet{bate03} and \citet{dan08,dan10}. As expected from linear theory, beyond $\approx 2h$ torque density drops dramatically, and the AMF rises only weakly, with $|x|$, which is evident from the flattening of AMF curves in Figure \ref{fig:amf}. This justifies the localization of the wave excitation to the immediate vicinity (within $\sim 2h$) of the planet, as described in \S\ref{sec:theory}.

To the best of our knowledge, no quantitative analytical description of how $F_H(x)$ should vary with distance from the planet exists in the literature, despite its potential importance for the gap opening problem. To fill this gap we first tried the following theoretical prescription $F_H^{\rm LR}(x)$ motivated by the calculations of GT80: at each $x$ we compute the lowest azimuthal wavenumber $\mu_{min}(x)$ for which the location of the Lindblad resonance $x_{L,m}<x$ using Eq. (\ref{eq:resonance-position}) and then calculate the AMF as the sum of all Fourier contributions $F_H(m)$ given by Eq. (\ref{eq:F_H}) that correspond to $m>(r_p/h)\mu_{min}(x)$. The result, which we denote by $F_H^{\rm LR}(x)$, is given by:
\ba
&& F_H^{\rm LR}(x)=\frac{4}{3}\frac{\Sigma_0(G \mplanet)^2}{c_s^2h}\left[2K_0(2/3)+K_1(2/3)\right]^2\int\limits_{\mu_{min}(x)}^\infty d\mu \mu^2 \frac{F_H(\mu)}{F_H^{\rm WKB}(\mu)},
\label{eq:F_H_phys}\\
&& \mu_{min}(x)=\frac{2}{3}\frac{h}{x},
\label{eq:mu_min}
\ea
where we adopt the torque cutoff function $F_H(\mu)/F_H^{\rm WKB}(\mu)$ calculated in GT80. Note that $F_H^{\rm LR}(x)$ is in fact independent of the adopted value of $r_p$ since its calculation can be rephrased fully in terms of $\mu$ rather than $m$.

This prescription essentially assumes that the AMF contribution $F_H(m)$ corresponding to a particular azimuthal wavenumber $m$ gets picked up by the density wave in a step-like fashion, solely at a {\it single discrete} location corresponding to the position of the $m$-th Lindblad resonance. Previously, this recipe was used in GT80 to compute the asymptotic behavior of the torque density in the limit $|x|\gtrsim h$, considering azimuthal harmonics with $1\lesssim m\lesssim r_{\rm p}/h$. Here we essentially extend this prescription to the case of arbitrary $m$ and $|x|$. Linear calculations by \citet{kor93} show that this approximation is not very accurate, and AMF contribution due to each potential harmonic is in fact accumulated by the wave over an extended range of $x$. Thus we should not expect $F_H^{\rm LR}(x)$ given by Eq. (\ref{eq:F_H_phys}) to accurately represent the real coordinate dependence of the AMF in the linear regime. Nevertheless, this prescription still provides a useful reference point and we plot it in Figure \ref{fig:amf}(a) (the dotted line, denoted by LR theory). We also plot in Figure \ref{fig:torque-density} the theoretical prescription for the torque density $dT_H^{\rm LR}(x)/dx$ obtained by differentiating Eq. (\ref{eq:F_H_phys}) with respect to $x$.

One can see from these Figures that the simple-minded theoretical prescription (\ref{eq:F_H_phys}) overestimates the torque density at small $x$ and underestimates it at larger $x$. While the areas under the numerical and theoretical curves in Figure \ref{fig:torque-density}, which represent the full accumulated AMF, are close to each other, the profiles of the two curves are quite different. As a result, $F_H^{\rm LR}(x)$ initially rises faster than the numerical $F_H(x)$ in Figure \ref{fig:amf}, but eventually the two asymptote to a similar final value. The small remaining difference is caused by both the intrinsic numerical inaccuracy and the fact that unlike the AMF theoretical curve the numerical AMF curve has been shifted to have a zero starting point.

As the next level of approximation to the behavior of the $F_H(x)$ and $T_H(x)$ we used the semi-analytical calculations of these quantities in the linear regime by Rafikov \& Petrovich (in preparation). The run of corresponding $T_H(x)$ is displayed in Fig. \ref{fig:amf} by the dash-dotted curve (the curve has also been shifted downward to cancel the non-zero starting value, and we use the label ``Linear theory'' to indicate Rafikov \& Petrovich's work in all the figures) and clearly demonstrates good agreement between the simulation results and the semi-analytical linear theory. Theoretical torque density based on this $T_H(x)$ is shown in Fig. \ref{fig:torque-density} and also agrees well with the numerical results, much better than the derivative of $F_H^{\rm LR}(x)$. This additionally emphasizes the point that assigning the planetary torque produced by a particular potential harmonic to a single location corresponding to the respective Lindblad resonance (as done in Eq. (\ref{eq:F_H_phys})) is not a very accurate procedure.

Finally, we also compare our numerical results with another quantity that can be computed in the linear approximation, namely the ``pseudo-AMF'' $f_J$ introduced in GR01 as
\ba
f_J(x)=\frac{2}{3}\frac{c_s^3}{\Omega\Sigma_0|x|}\int\limits_{-\infty}^\infty\left(\delta\Sigma\right)^2 dy.
\label{eq:pseudo}
\ea 
This quantity reduces to $F_H(x)$ at large separations from the planet $|x|\gtrsim h$, where $\delta\Sigma$ is determined solely by the density wave, propagating away from the planet. However, for $|x|\lesssim h$ the pseudo-AMF is strongly affected by the presence of the aforementioned atmosphere around the planet (see \S\ref{sec:density}), which is not a propagating density perturbation and cannot be associated with the density wave (this explains the prefix ``pseudo-''). 

We compute the theoretical value $f_J(x)$ in the linear approximation using unpublished data from GR01 and also derive $f_J(x)$ from our numerical data using the definition (\ref{eq:pseudo}). The comparison between the two is shown in Figure \ref{fig:flux2} for very small $\mplanet=3.7\times10^{-3}\Mth$. One can indeed see that at small $|x|$ the pseudo-AMF {\it increases} with decreasing $|x|$, analogous to $\delta\Sigma$ in Figure \ref{fig:density-profile}a, which is explained by the presence of the atmosphere accumulated in the planetary potential well. Comparison with Figure \ref{fig:amf} also shows that $f_J(x)$ indeed reduces to $F_H(x)$ at large $|x|$. Most importantly, the agreement between the theoretical $f_J(x)$ derived from linear theory and the numerical $f_J(x)$ is very good in the whole range of $x$ explored, which provides additional verification of linear theory of the planetary density wave evolution.


\subsection{Sensitivity to EOS}
\label{sec:EOS}

In linear theory the dynamics of the density wave is independent of the adopted equation of state (EOS), and depends only on adiabatic sound speed of the gas $\cs$. To check this property we have run several test simulations with two different EOS: isothermal EOS and EOS with $\gamma=5/3$, while keeping $\cs$ the same. The results do not show any significant difference between the runs with different EOS in linear stage thus confirming theoretical expectations. However, in the nonlinear stage a particular form of the EOS does become important and this is investigated further in Paper II.


\subsection{Emergence of the nonlinear effects}
\label{sec:nonlinear}

We now discuss the nonlinear effects that arise in our simulations even prior to the appearance of the shock, i.e. at $|x|\lesssim \ls$. Nonlinear distortion of the wave profile is unavoidable even for the very low amplitude density waves and even during the mostly linear phase of their evolution. The rate of accumulation of the nonlinear effects (which eventually become strong and lead to shock formation) scales with planetary mass and is lower for low $\mplanet$. 

\begin{figure}[tb]
\begin{center}
\epsscale{0.45} \plotone{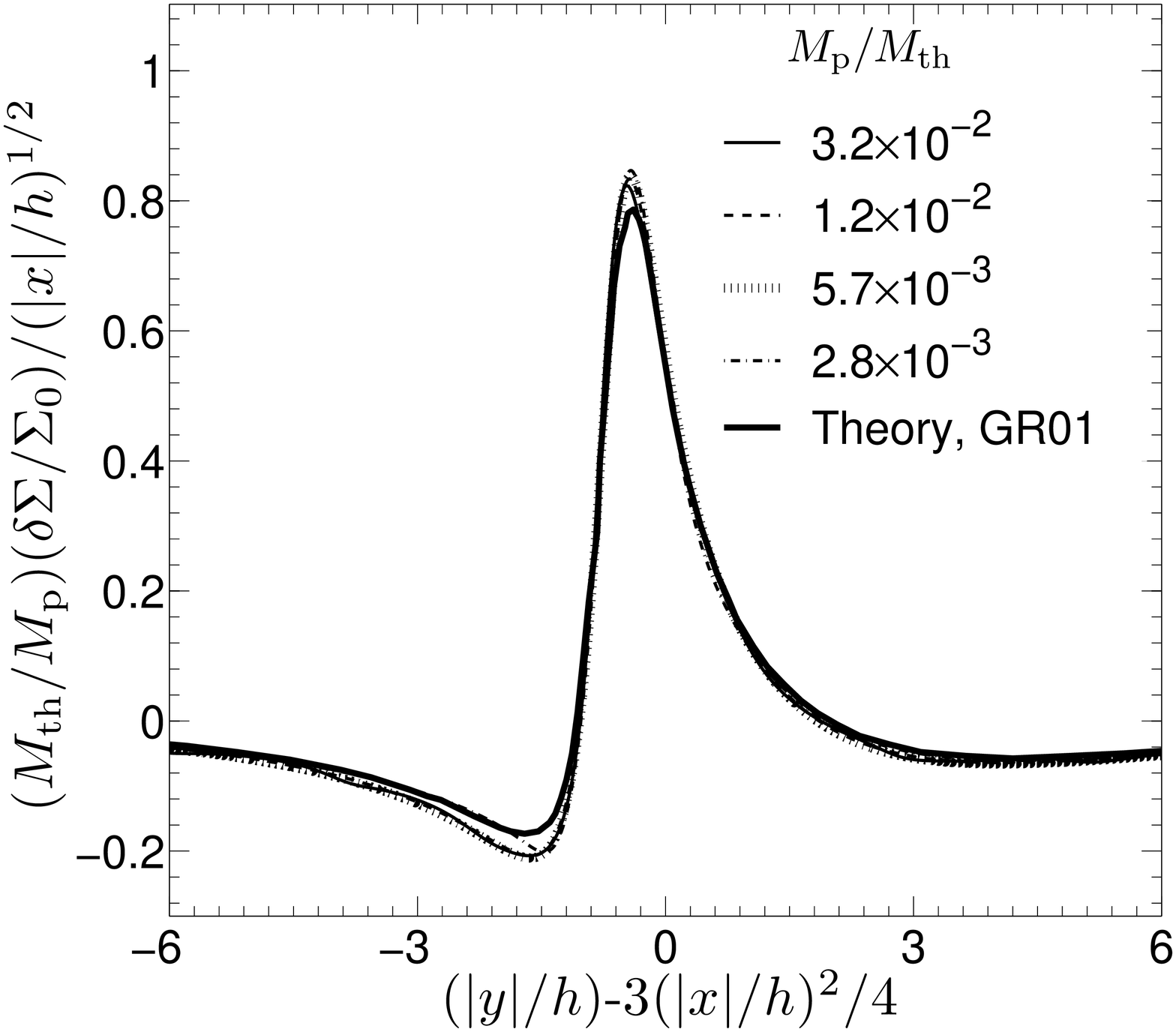} \plotone{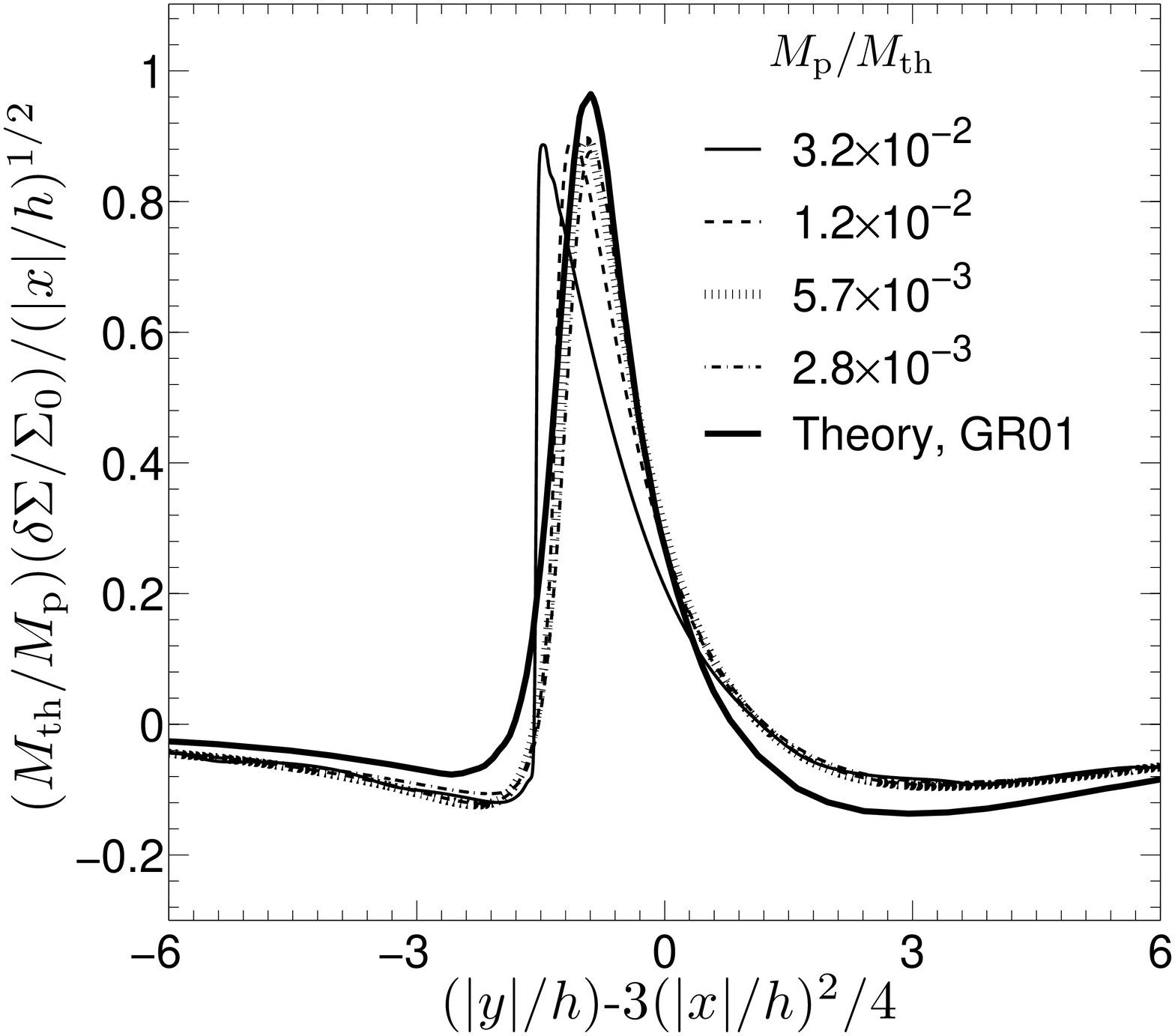}
\end{center}
\figcaption{Azimuthal (normalized) profiles of the density perturbation for different $\mplanet$ (labeled in panels) at two radial distances from the planet: (a) $x=1.33h$ and (b) $x=4h$. Analytical density profile from GR01 is shown by thick solid curve. See text for details.
\label{fig:mp}}
\end{figure}

\begin{figure}[tb]
\begin{center}
\epsscale{0.45} \plotone{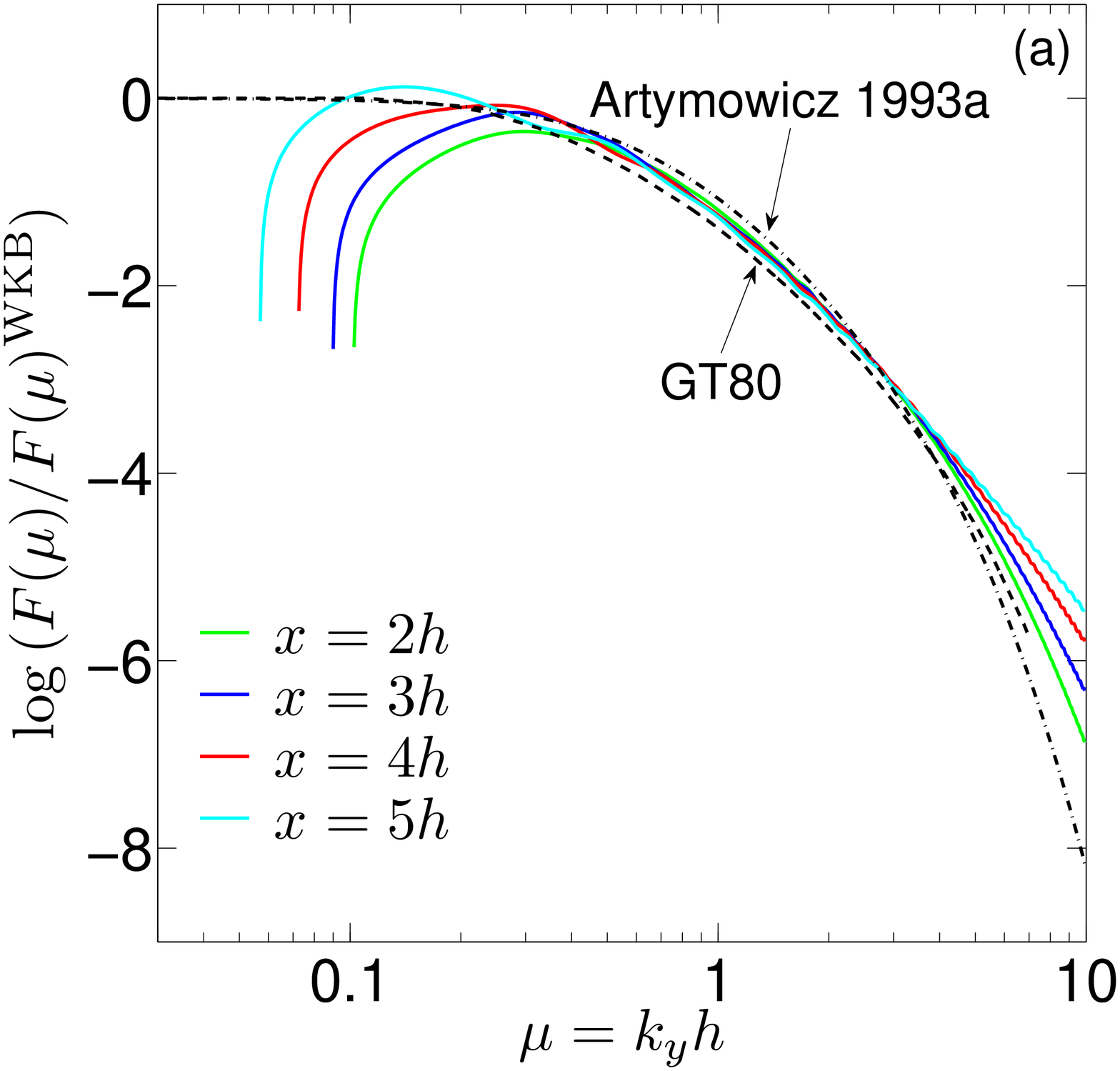}\plotone{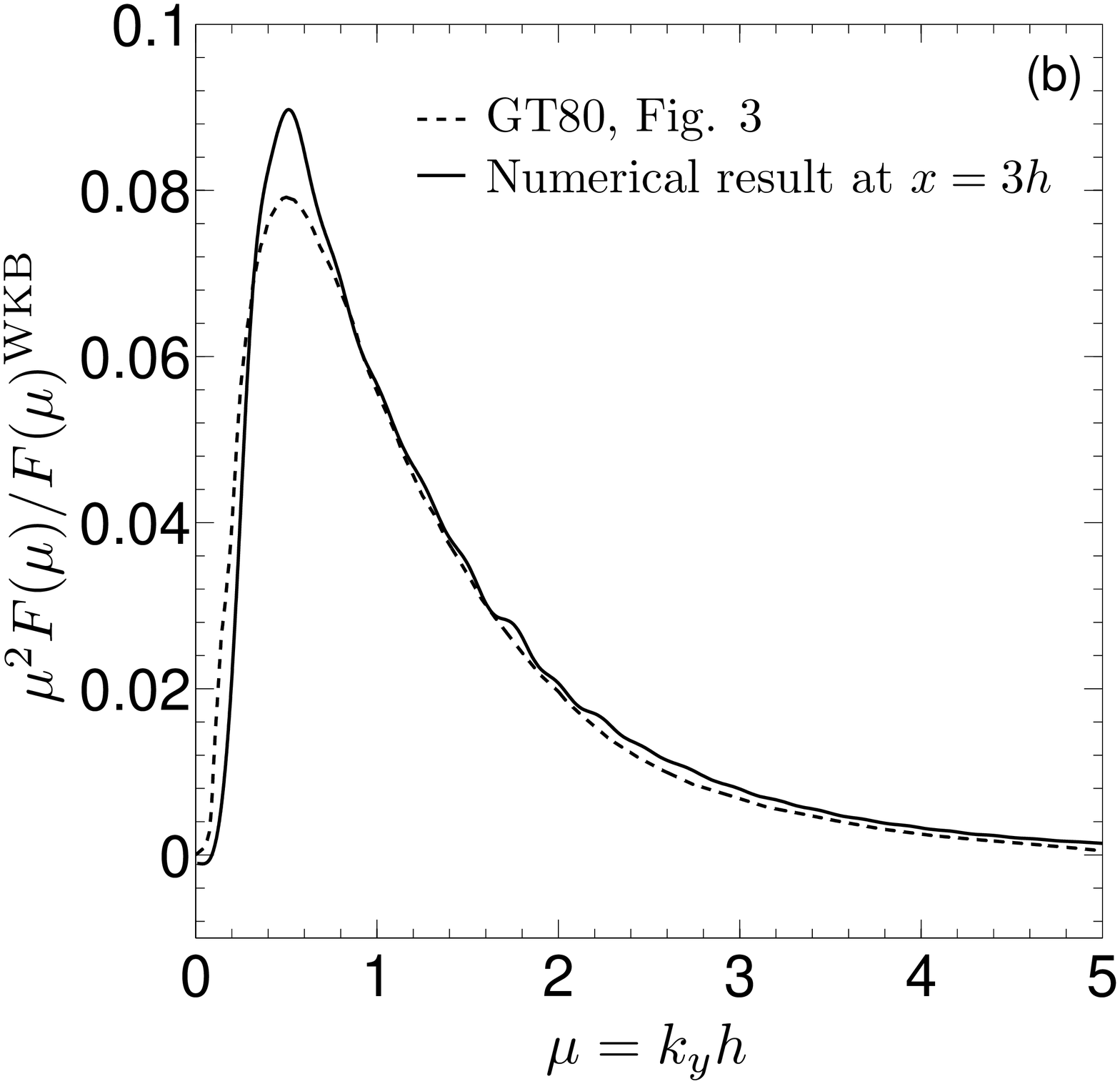}
\end{center}
\figcaption{Spectral decomposition of the angular momentum flux $F_H(x)$ carried by the wave in azimuthal Fourier harmonics. Numerical results are based on isothermal simulation for $\mplanet=1.2\times 10^{-2}\Mth$ (corresponding to $0.14$ M$_\oplus$ at 1 AU and $\ls\approx 5h$) with $128/h$ resolution, $\rs=h/16$. (a) Ratio $F_{H,k}(x)/F_{H,k}^{WKB}$ (the so-called {\it torque cutoff function}) of the numerical Fourier spectrum of the AMF to the analytical WKB calculation of GT80 as a function of $k_yh$, plotted at several values of $x$. For comparison we show analogous quantity (in the limit $|x|\to \infty$) computed by GT80 (see their Figure 2) and \citet[their Eq. 25]{art93a}. The origin of excess power at high $k_y$ is discussed in \S\ref{sec:nonlinear}. (b) The same quantity multiplied by $\mu^2$ (using the $x=3h$ curve) and plotted on linear scale to facilitate comparison with Figure 3 of GT80.
\label{fig:amf-decay}}
\end{figure}

\begin{figure}[tb]
\begin{center}
\epsscale{0.45} \plotone{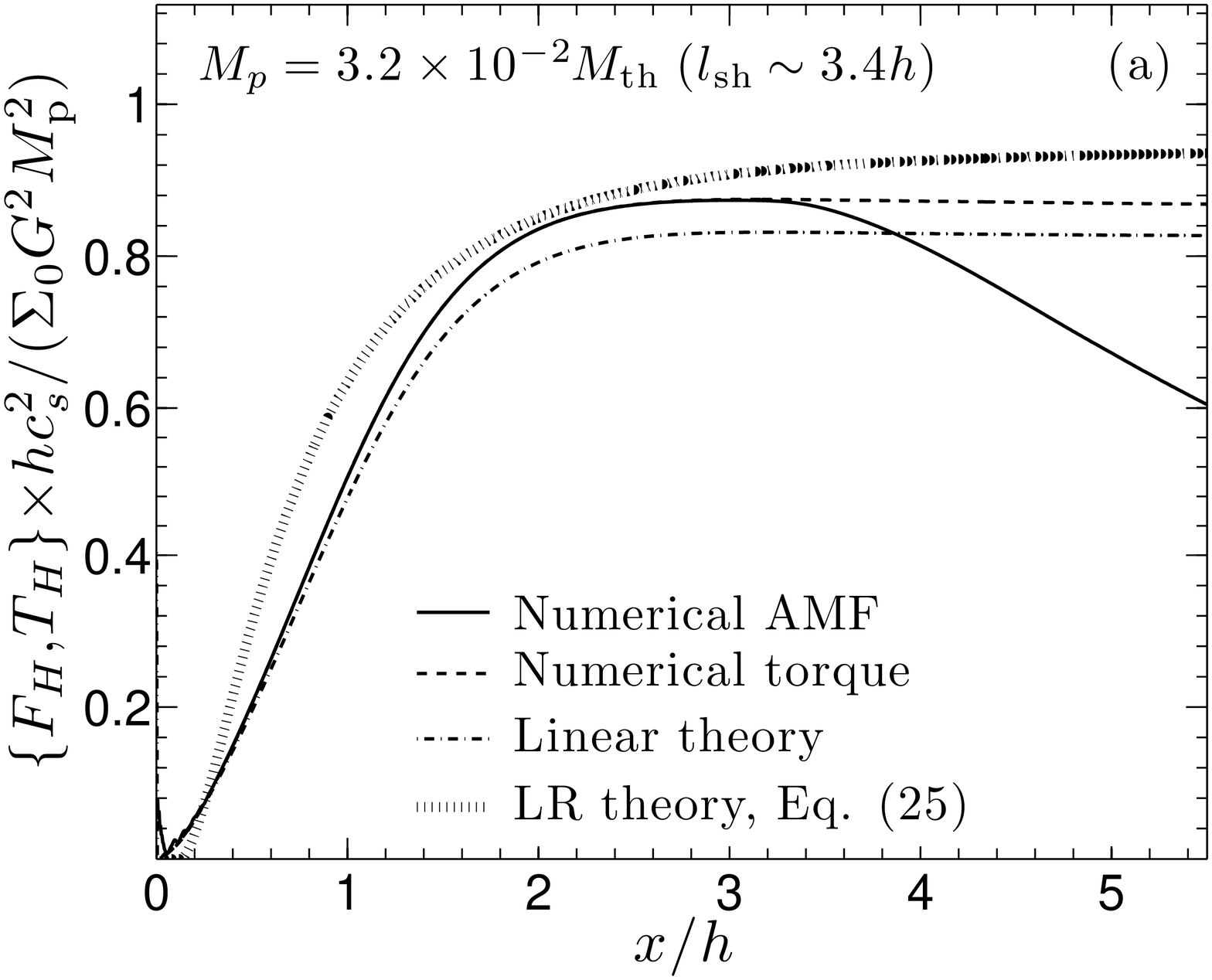} \plotone{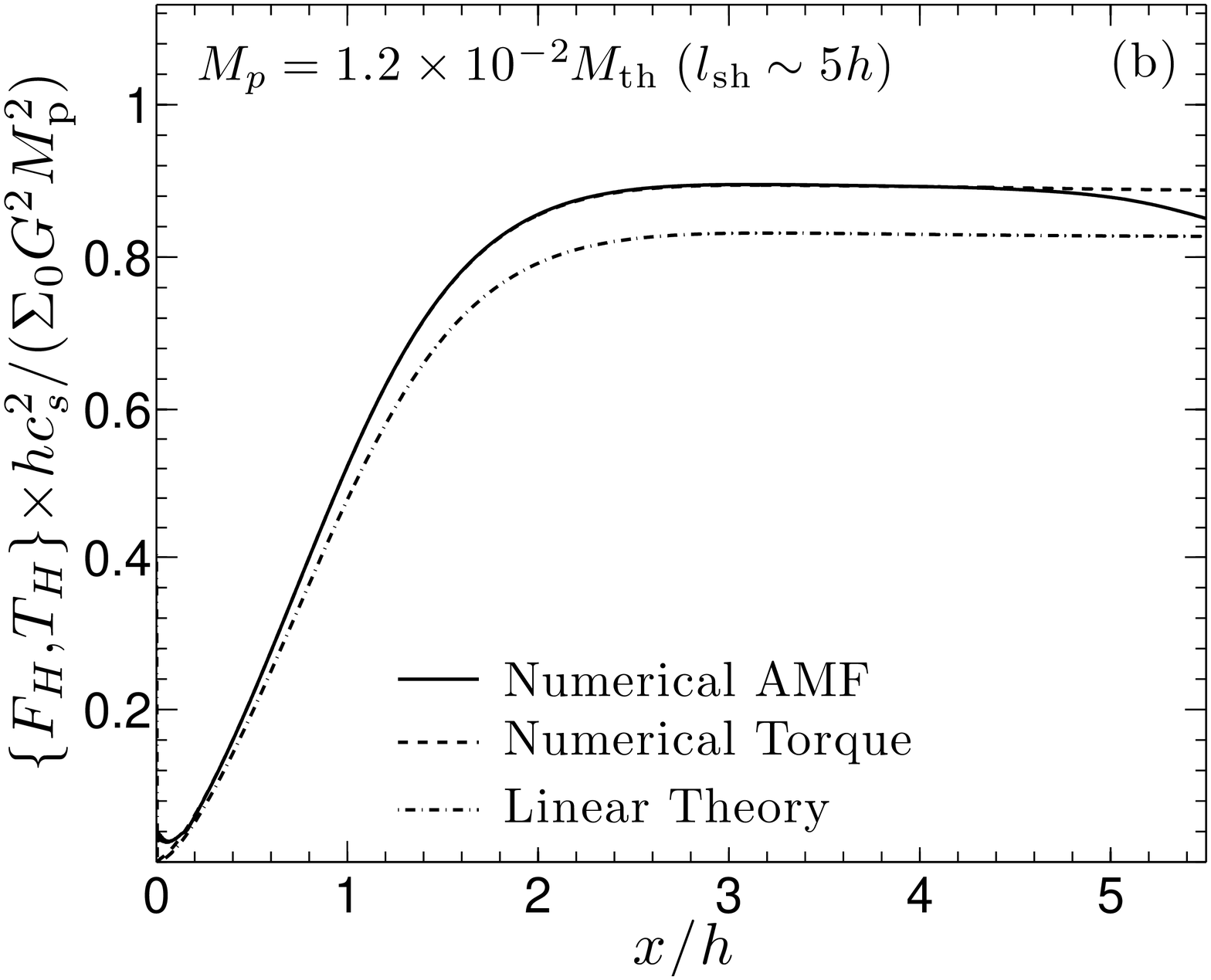} \plotone{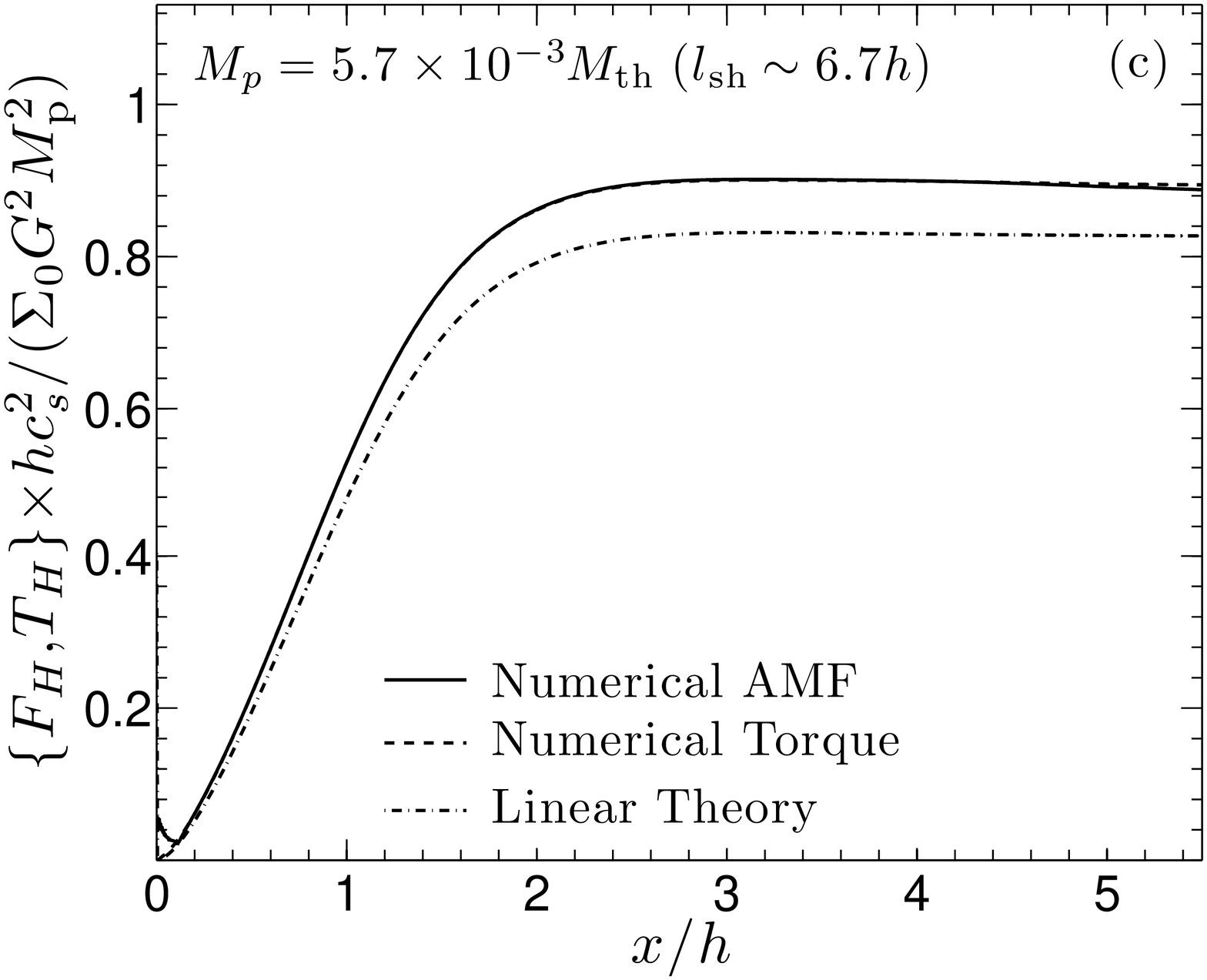} \plotone{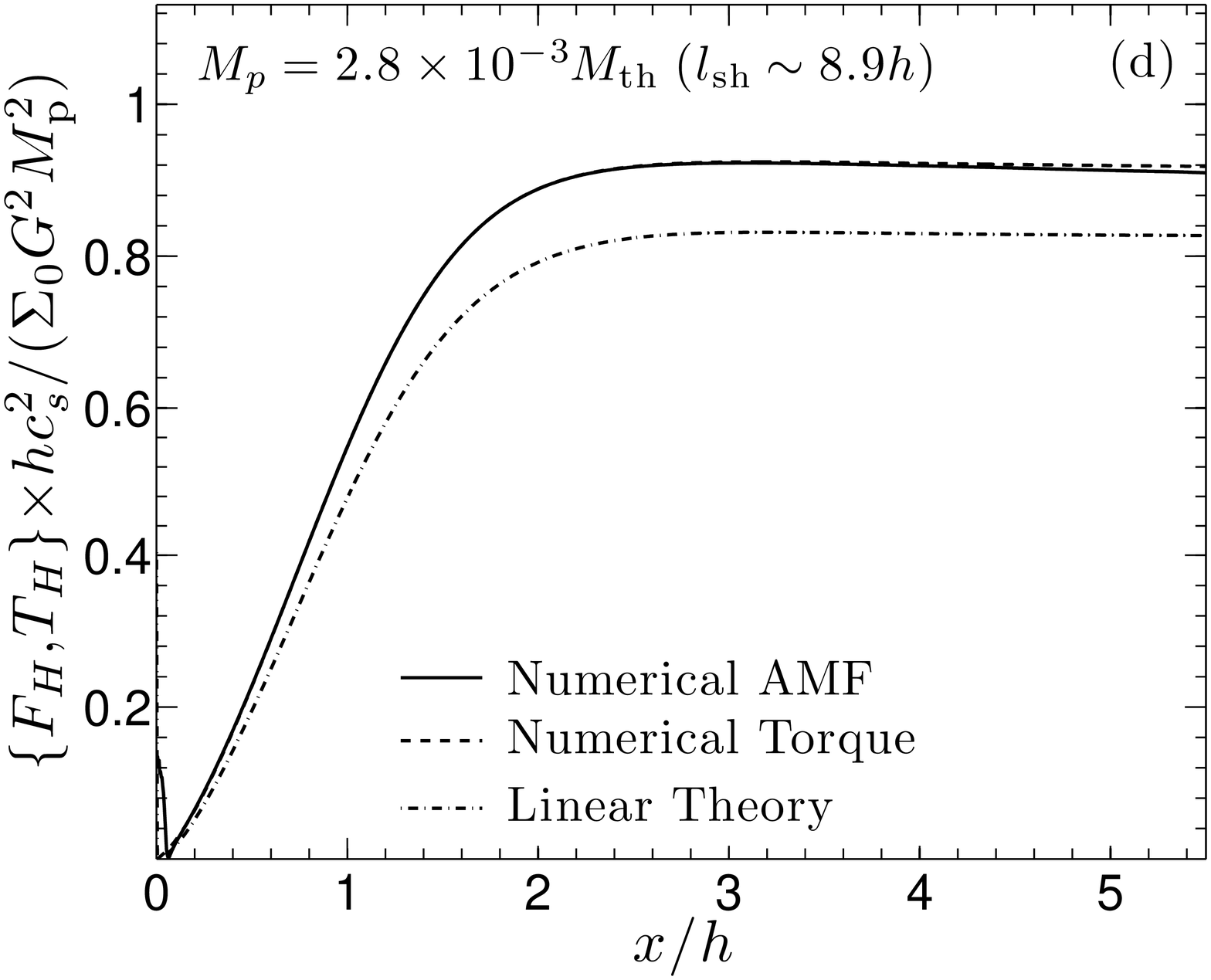}
\end{center}
\figcaption{Integrated torque $T_H(x)$ and the angular momentum flux carried by the wave $F_H(x)$ (shifted vertically so that $F_H(0)=0$ to simplify comparison with $T_H(x)$) as a function of $x$. Different panels correspond to different values of $\mplanet$ (labeled on the plot, the same as in Figure \ref{fig:mp}). Analytical prescription (\ref{eq:F_H_phys}) for $T_H(x)$ motivated by the results of GT80 is shown in panel (a) by the dotted curve. The theoretical prediction based on the linear semi-analytical calculation (Rafikov and Petrovich, in preparation) is shown by the dash-dotted line in all panels. Both $F_H(x)$ and $T_H(x)$ are scaled by $\Sigma_0(G\mplanet)^2/(h\cs^2)$, which should make their shapes independent of $\mplanet$ in the framework of linear theory.
\label{fig:amf}}
\end{figure}

\begin{figure}[tb]
\begin{center}
\epsscale{0.75} \plotone{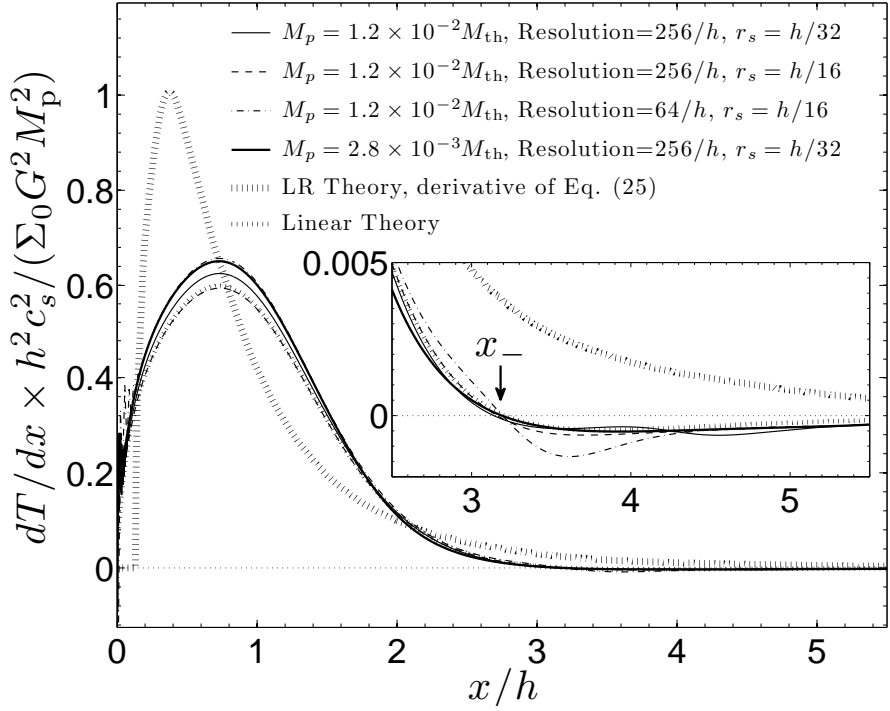}
\end{center}
\figcaption{Torque density $dT_H(x)/dx$ (Eq. \ref{eq:torque-density}) as a function of $x$ computed for two different values of $\mplanet$ and different numerical parameters. The thick dotted curve shows the derivative of Eq. (\ref{eq:F_H_phys}) motivated by the results of GT80, and the thin dotted curve shows the linear theory result based on the semi-analytical calculation by Rafikov and Petrovich (in preparation). Torque density is scaled by $\Sigma_0(G\mplanet)^2/(h^2\cs^2)$ removing the dependence on $\mplanet$ in linear approximation. For $\mplanet=1.2\times 10^{-2}\Mth$ we show results from three simulations with different adopted resolution or $\rs$. An inset zooms in on a region near $x=3h$, where we find the torque density changing sign. The location ($x_-=3.2h$, shown by arrow) at which this happens is found to be insensitive to variations of simulation parameters or $\mplanet$ ($\mplanet=2.8\times 10^{-3}\Mth$ was also explored). 
\label{fig:torque-density}}
\end{figure}

\begin{figure}[tb]
\begin{center}
\epsscale{0.45} \plotone{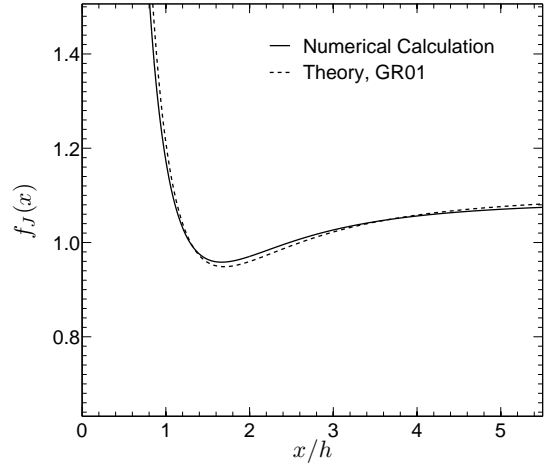}
\end{center}
\figcaption{``Pseudo-AMF'' defined by Eq. (\ref{eq:pseudo}) from our numerical simulations ({\it solid curve}) and semi-analytical linear calculations of GR01 ({\it dashed curve}). The agreement between the two is very good in the linear regime (for $\mplanet=3.7\times 10^{-3}\Mth$ used in simulation wave shocks at $\ls\approx 8h$).
\label{fig:flux2}}
\end{figure}

This point is illustrated in Figure \ref{fig:mp}b, which is very similar to Figure \ref{fig:mp}a and presents azimuthal density cuts through the wake for different values of $\mplanet$, but now recorded further out from the planet, at $x=4h$. While in Figure \ref{fig:mp}a, at $x=1.33h$, the density profiles for all $\mplanet$ are essentially overlapping and agree with the analytical profile from GR01, the situation at $x=4h$ is quite different. 

At this location the (normalized) wake profiles for the two smallest masses, $\mplanet=2.8\times10^{-3}\Mth$ (corresponding to $l_s\approx8.9h$) and $\mplanet=5.7\times10^{-3}\Mth$ ($l_s\approx6.7h$), still agree with each other and the semi-analytical linear calculation from GR01 quite well. However, already for $\mplanet=1.2\times10^{-2}\Mth$ ($l_s\approx5h$) one can see a noticeable distortion of the profile  compared to the linear solution, with the leading edge (at the left) becoming {\it steeper} and the profile peak {\it shifting} to the left, towards incoming fluid. The wave for even higher $\mplanet=3.2\times10^{-2}\Mth$ ($l_s\approx 3.4h$) has already shocked at $x=4h$, which is clearly reflected in its density profile: its leading edge is essentially vertical, exhibiting a discontinuity in the fluid variables across the shock. Thus, the role of the nonlinear effects, which is measured at a given $x$ both by the slope of the leading edge of the profile and the shift of its peak relative to the linear solution, progressively increases with the planet mass.

Nonlinear effects also manifest themselves in a variety of other, more indirect ways. In particular, they explain the evolution of the torque cutoff function in Fourier space and its deviation from theoretical predictions at high $k_yh\gtrsim 5$ (see \S\ref{sec:Fourier}). Figure \ref{fig:amf-decay}a clearly shows a growing amount of excess power at the high values of $k_yh$ as the wave propagates to larger $x$. This behavior is naturally accounted for by the nonlinear wave steepening, which causes the transfer of the AMF power in Fourier space from the low-$k_y$ to high-$k_y$ harmonics.  

Nonlinearity also affects the AMF behavior in physical space. Close inspection of Figure \ref{fig:amf} reveals that the numerical torque and AMF curves tend to asymptote to lower levels for higher $\mplanet$, when nonlinearity is stronger. The difference in asymptotic values at large $x$ between the largest and the smallest planet cases is about $5\%$. Variation of $\mplanet$ also affects the height of $dT_H(x)/dx$ curve at smaller $x\approx h$, increasing the peak value of $dT_H(x)/dx$ for lower $\mplanet$, as shown in Figure \ref{fig:torque-density}. This explains the dependence of normalized AMF and $T_H(x)$ curves on $\mplanet$ in Figure \ref{fig:amf}: the area under the differential torque curve is slightly (by $\approx 3\%$) higher for $\mplanet=2.8\times 10^{-3}\Mth$ (the think solid line) than for $\mplanet=1.2\times 10^{-2}\Mth$ (the thin solid line), explaining the difference seen between the asymptotic values of integrated torque $T_H(x)$ in Figures \ref{fig:amf}b and \ref{fig:amf}d. Spurious dissipation related to numerical viscosity cannot be blamed for this effect since its effect on the AMF should be independent of $\mplanet$.

We suggest that this behavior may be explained (at least partly) by the stronger nonlinear evolution of the wave profile for higher $\mplanet$. According to Eq. (\ref{eq:torque-density}) nonlinear distortion of $\delta\Sigma$ away from the theoretically expected-value should have an effect (growing with $\mplanet$) on the calculation of the torque density $dT_H(x)/dx$. Since for $\mplanet$ not too close to $\Mth$ this profile distortion becomes significant only far from the planet, where the planetary potential is weak, the impact of this nonlinear torque modification on the AMF curves is not very dramatic, but it is definitely non-zero. We also note that one expects simulations with smaller $\mplanet$ to exhibit better agreement with the theory since they are less affected by the nonlinear effects. However in reality this is not always the case because the optimal set of numerical parameters for achieving the best possible numerical result for different $\mplanet$ is different (see \S \ref{sec:parameters}).


\subsection{``Negative torque'' phenomenon}
\label{sec:neg_torque}

Close inspection of Figure \ref{fig:torque-density} reveals an interesting feature in the behavior of the torque density at large $x$. The inset in this Figure clearly shows that $dT_H(x)/dx$ becomes {\it negative} beyond $x_-\approx 3.2h$, which is at odds with linear theory since according to GT80 $dT_H(x)/dx$ should not change sign at any non-zero $x$.  The contribution of this negative torque density to the total integrated torque $T_H$ is rather small (under one per cent level), but its existence presents a challenge to the results of GT80.  Further investigation reveals that this ``negative torque'' phenomenon is present also in previous lower resolution numerical calculations of \citet[Figure 12]{bate03} and \citet[Figure 7]{dan08}. 

To understand the nature of this effect we varied the numerical parameters of our simulations. For a given $\mplanet=1.2\times 10^{-2}\Mth$ we tried varying softening length $\rs$ from $h/16$ to $h/32$ at fixed resolution of $256/h$, and then varied resolution from $h/64$ to $h/256$ for a fixed $\rs=h/16$. As Figure \ref{fig:torque-density} shows this does not make the negative torque go away and $dT_H(x)/dx$ still changes sign at the same value of $x_-$. Even more interestingly, changing the planetary mass $\mplanet$ to lower $\mplanet=2.8\times 10^{-3}\Mth$ while keeping everything else the same (resolution of $256/h$ and $\rs=h/32$) also does not affect the position of $x_-$, as the same Figure clearly shows. Analogously, experiments with different sizes of the simulation box and different prescriptions of gravitational potential show that the existence of negative torque and the position of $x_-$ are independent of these parameters as well. Moreover, simulations of \citet[Figure 12]{dan08} suggest that the negative torque phenomenon shows up only at {\it low} planetary masses ($\mplanet\lesssim 0.03$ M$_J\approx 10$ M$_\oplus$ in their case), which makes possible explanations based on nonlinear effects unlikely. 

This point is confirmed in Rafikov \& Petrovich (in preparation) who demonstrate that the sign change of the torque density at large separation from the planet is in fact a purely linear effect. It can be understood by carefully accounting for the complex spatial structure of the torque density produced by each azimuthal harmonic of the planetary potential, which goes beyond the simple calculation presented in GT80.


\section{Sensitivity of results to numerical parameters}
\label{sec:parameters}


We now explore how the results presented in previous section and their agreement with analytical theory are affected by the variation of purely numerical parameters in our simulations. Table \ref{tab:parameter} lists the numerical parameters that we varied and the values we explored. Values corresponding to our standard case are indicated in boldface.

\begin{deluxetable}{lc}
\tabletypesize{\scriptsize}
\tablewidth{0pt}
\tablecaption{Parameter space of the simulations}
\tablehead{
Parameters\tablenotemark{a} & Range
}
\startdata
Riemann solver (flux function) used & {\bf Roe}, HLLC \\
Order of accuracy & {\tt 2}, {\tt {\bf3}{\bf c}}, {\tt 3p} \\
Boundary conditions in $y$ & Outflow, {\bf Inflow/Outflow} \\
Resolution of the simulation (cells per $h$) & 64, 128, {\bf 256} \\
Planetary potential\footnote{see \S \ref{sec:code}} & $\phip^{(2)}$, $\bf \phip^{\bf(4)}$, $\phip^{(6)}$ \\
Softening length & 1/8, 1/16, {\bf 1/32} \\
Equation of states of the fluid & {$\bf \gamma=1$}, $\gamma=5/3$ \\
Mass of the planet $\mplanet/\Mth$ & $3.2\times10^{-2}$, $1.2\times10^{-2}$, $5.7\times10^{-3}$, $3.7\times10^{-3}$, $2.8\times10^{-3}$ \\
\enddata
\tablenotetext{a}{See Section~\ref{sec:parameters} for details.}
\label{tab:parameter}
\end{deluxetable}


\subsection{Numerical solver and order of accuracy}
\label{sec:solver}

In our simulations we compare two different Riemann solvers --- Roe's linearized solver (\citealt{roe81}) and HLLC \citep{tor99} --- with three different algorithms for the spatial reconstruction step \citep{sto08}: second order with limiting in the characteristic variables (denoted {\tt 2} in this work), which is the predominant choice in literature in this field, third order with limiting in either the characteristic variables ({\tt 3c}), or in the primitive variables ({\tt 3p}). We find that Roe and HLLC solvers yield nearly identical results both in terms of the density profile and in terms of $F_H(x)$ and $T_H(x)$ (differences are at the $0.1\%$ level).

The effect of using a different order of accuracy on the profile of the density wave is shown in Figure \ref{fig:density-variety}a, where we plot azimuthal density cuts obtained with different numerical settings at $x=1.33h$ (together with semi-analytical profile computed by GR01 in the framework of linear theory) and $x=4h$. One can see that density perturbation is rather insensitive to variation of the order of accuracy.

The sensitivity of the AMF calculation to varying order of accuracy is illustrated in Figure \ref{fig:amf-variety}a, where the linear theory curve based on the semi-analytical calculation by Rafikov and Petrovich (in preparation) is overplotted for reference. One can see that the higher order of accuracy ({\tt 3c} and {\tt 3p}) results in a slightly lower asymptotic value of torque and AMF, and the difference in the asymptotic value for the two between the {\tt 2}  case and the {\tt 3p} case is about $5\%$. Furthermore, {\tt 3c} and {\tt 3p} cases agree with each other on the position where the AMF curve starts to deviate from the torque, $\sim4h$, which is still somewhat smaller than the theoretical prediction $\ls\approx5h$ in this case due to the relatively low resolution. However, the second order of accuracy moves this point inward, which means that the angular momentum carried by the wave starts to dissipate earlier in this case. In particular, we find that using a second order accuracy solver compared with the third order accuracy solver has similar effect to decreasing the resolution by a factor of 2 (in term of advancing the displacement of the AMF-torque separation point, see \S \ref{sec:resolution} and Figure~\ref{fig:amf-variety}b).

Finally, we also note that calculations with {\tt 3p} order of accuracy produce considerably noisier velocity field (this explains the noisy corresponding AMF curve in Figure \ref{fig:amf-variety}a) than the other two cases explored. This difference is important for calculation of the velocity-based variables such as potential vorticity, which is used in Paper II as the means of shock detection. For these reasons in our current simulations we use Roe solver with {\tt 3c} reconstruction.


\subsection{Resolution}
\label{sec:resolution}

In Figure \ref{fig:density-variety}b we show the effect of varying resolution of our simulations on the density profile. In general, increasing resolution from $64/h$ to $256/h$ improves the agreement with linear theory, but only slightly. Lower resolution simulations overestimate the amplitude of $\delta\Sigma$ by just several per cent compared to $256/h$ simulations. Thus, for the study of the density wave profile our simulations essentially reach convergence in terms of resolution already at $64/h$. 

In Figure \ref{fig:amf-variety}b we show that the AMF and torque calculations are more sensitive to resolution, especially when the nonlinear effects become important. Increasing resolution causes the asymptotic values of $F_H(x)$ and $T_H(x)$ to decrease, and the difference in the asymptotic value for the two between the $256/h$ case and the $64/h$ case is about $10\%$. The key factor that clearly shows the downside of low resolution is the location, at which the numerical AMF curve deviates from the numerical torque calculation. Calculations shown in Figure \ref{fig:amf-variety} use $\mplanet=1.2\times10^{-2}\Mth$, which according to Eq. (\ref{eq:ls}) corresponds to shocking length $\ls\approx 5h$. In our highest resolution simulations ($256/h$), the AMF curve starts to deviate from the accumulated torque curve precisely at $\ls\approx 5h$, as expected from theory. But as we reduce resolution, the radial separation at which $F_H(x)$ starts departing from $T_H(x)$ moves closer to the planet violating the agreement with theory. In $64/h$ simulations such departure starts already at $3.2h$, considerably closer to the planet than the nominal shock location $\ls$. This behavior is caused by the higher level of numerical viscosity arising at low resolution, which leads to the dissipation of angular momentum carried by the wave prior to shock formation (we verified this point by conducting some runs with explicit viscosity in paper II, which show behavior qualitatively similar to our low resolution runs here). Thus, high resolution is crucial to correctly capture the details of the nonlinear wave evolution. We elaborate on this issue in Paper II.


\subsection{Planetary potential and softening length $\rs$}
\label{sec:phip+rs}

The rate at which a given smoothed potential converges to $\Phi_K$ is important for the problem of density wave generation. Indeed, the amplitude and spatial distribution of fluid perturbation is in the end determined solely by the potential behavior. As a result, if $\Phi$ is substantially different from $\Phi_K$ in the region where most of the torque is exerted by the planet, i.e. at $|x|\approx 0.2-2h$, see Figure \ref{fig:torque-density}, then one can expect significant spurious effects modifying the density wave properties.  

A specific form of the potential can have a two-fold effect on the properties of the density wave excited by the planet. First, there is a deviation of the potential from Newtonian, which directly affects the wave excitation at a given separation from the planet. But in addition, different potentials result in different structures of the quasi-static atmospheres (see \S\ref{sec:density}) that get collected in the planetary potential well. This has an effect on the pressure distribution in the vicinity of the planet and can also affect wave excitation by modifying the dispersion relation and displacing the effective positions of Lindblad resonances. In practice disentangling these two effects based on the results of simulations may be non-trivial. 

\begin{figure}[tb]
\begin{center}
\epsscale{0.45} \plotone{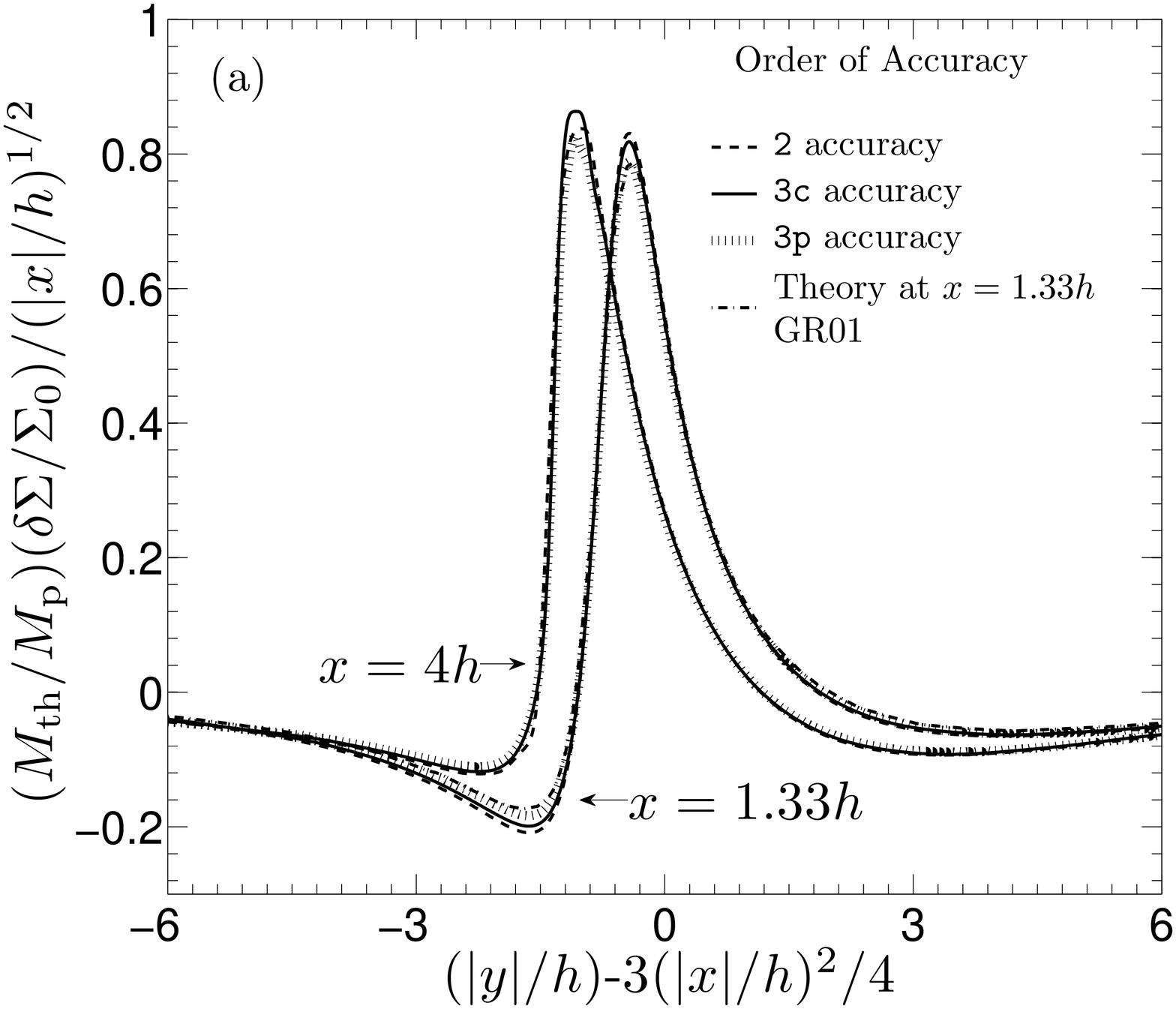} \plotone{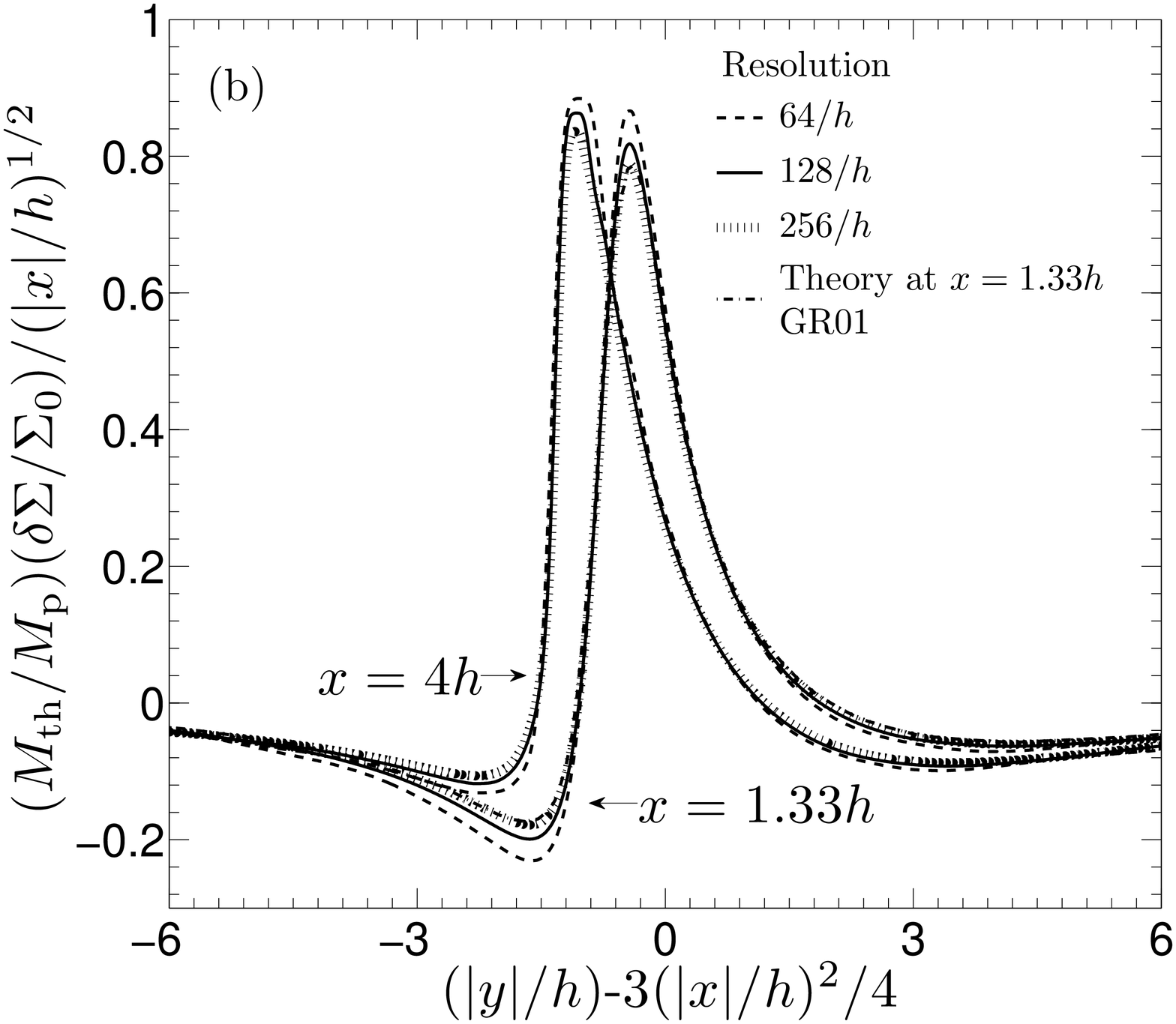} \plotone{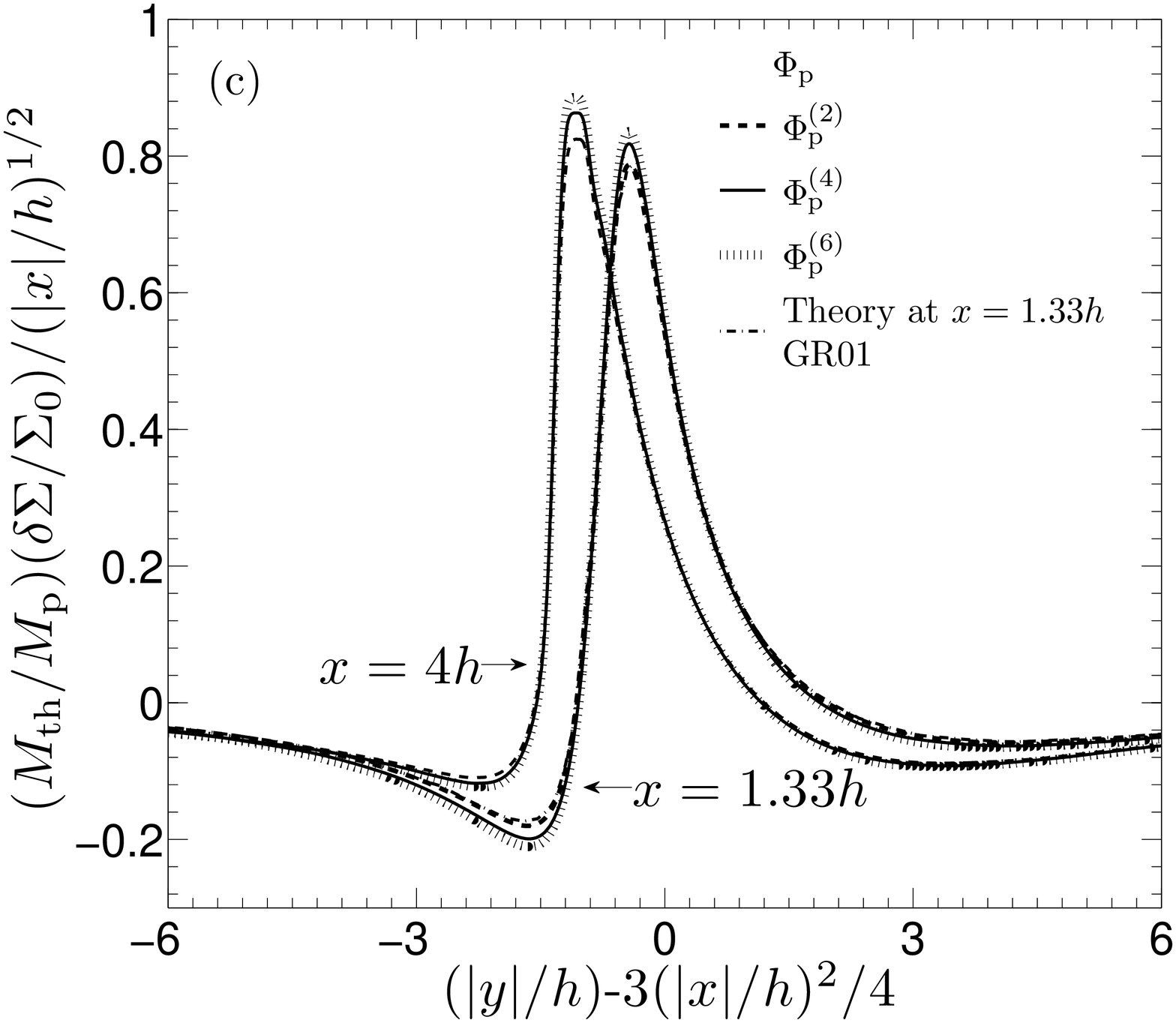} \plotone{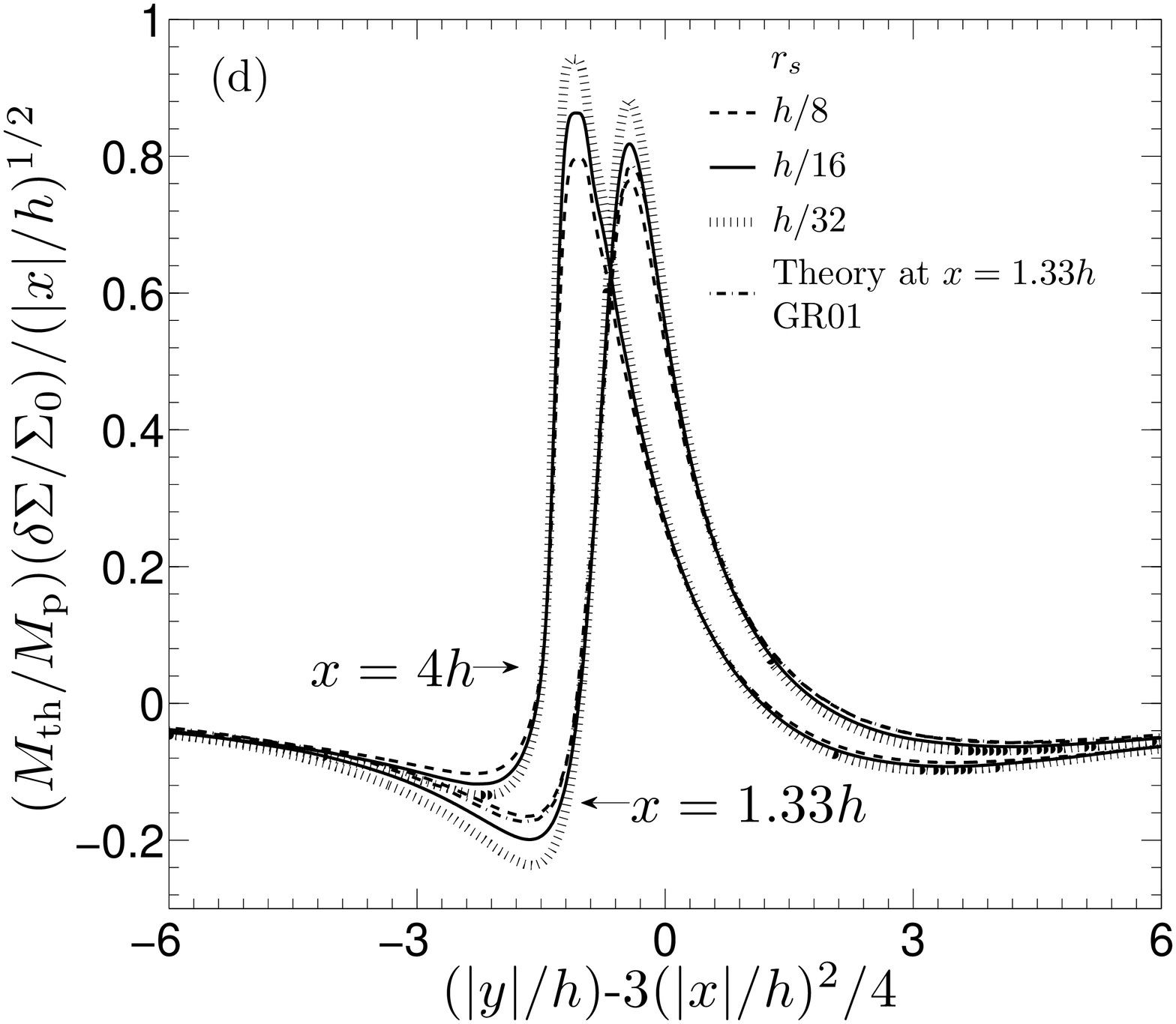}
\end{center}
\figcaption{Results of the numerical convergence study. Different panels show results of simulations (normalized azimuthal density cuts, analogous to those shown in Figure \ref{fig:mp}, at $x=1.33h$ and $x=4h$) in which a single numerical parameter was varied: (a) order of accuracy for the solver, (b) resolution, (c) softened planetary potential, (d) softening length of the potential. The standard set of numerical parameters is $\mplanet=1.2\times 10^{-2}\Mth$, $\phip^{(4)}$ potential, $\rs=h/16$, $128/h$ resolution, and {\tt 3c} accuracy. Results of the simulation with this set are shown by solid line in all panels. The semi-analytical density profile from linear calculations of GR01 is shown only at $x=1.33h$ to guide the eye.
\label{fig:density-variety}}
\end{figure}

\begin{figure}[tb]
\begin{center}
\epsscale{0.45} \plotone{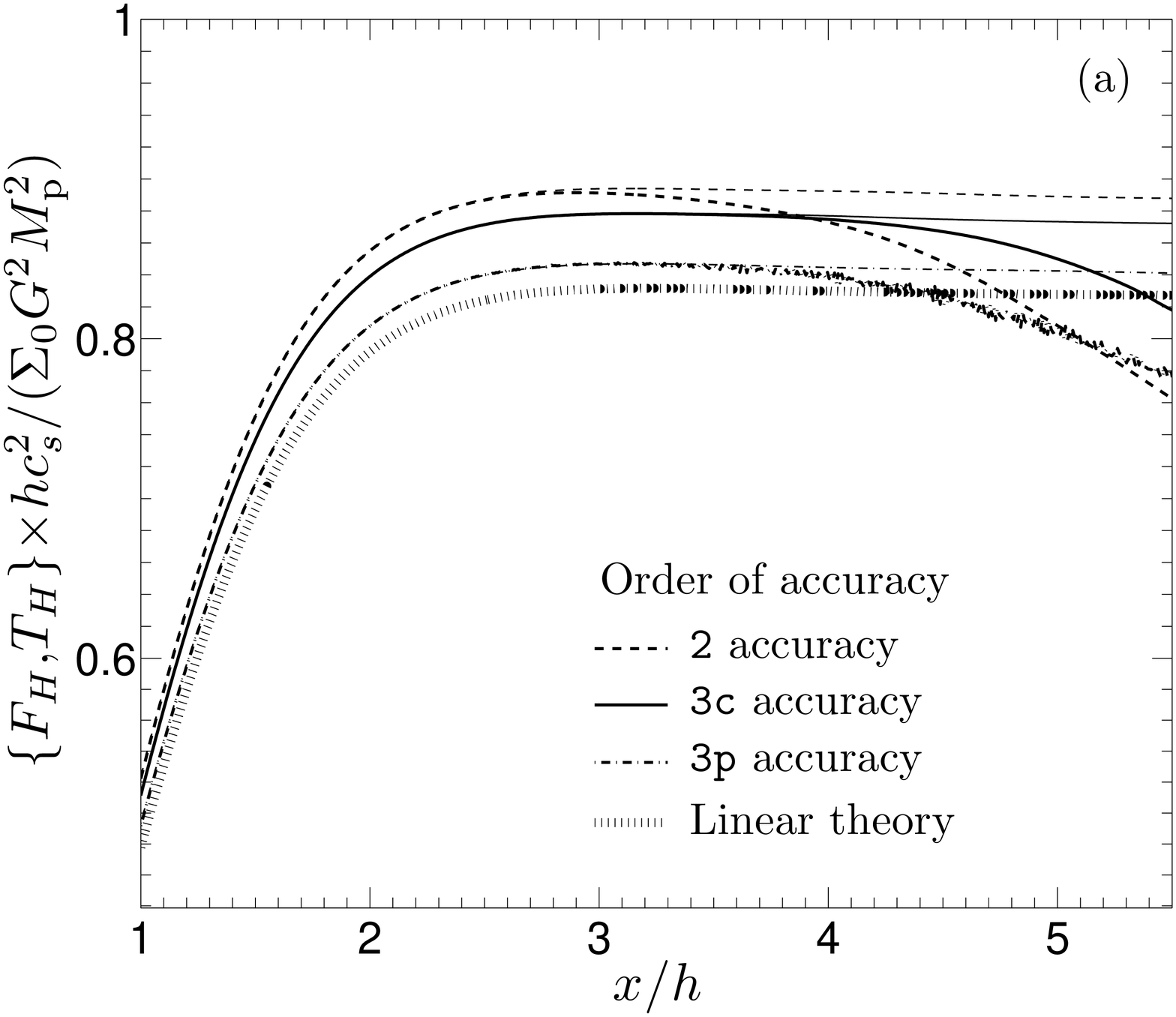} \plotone{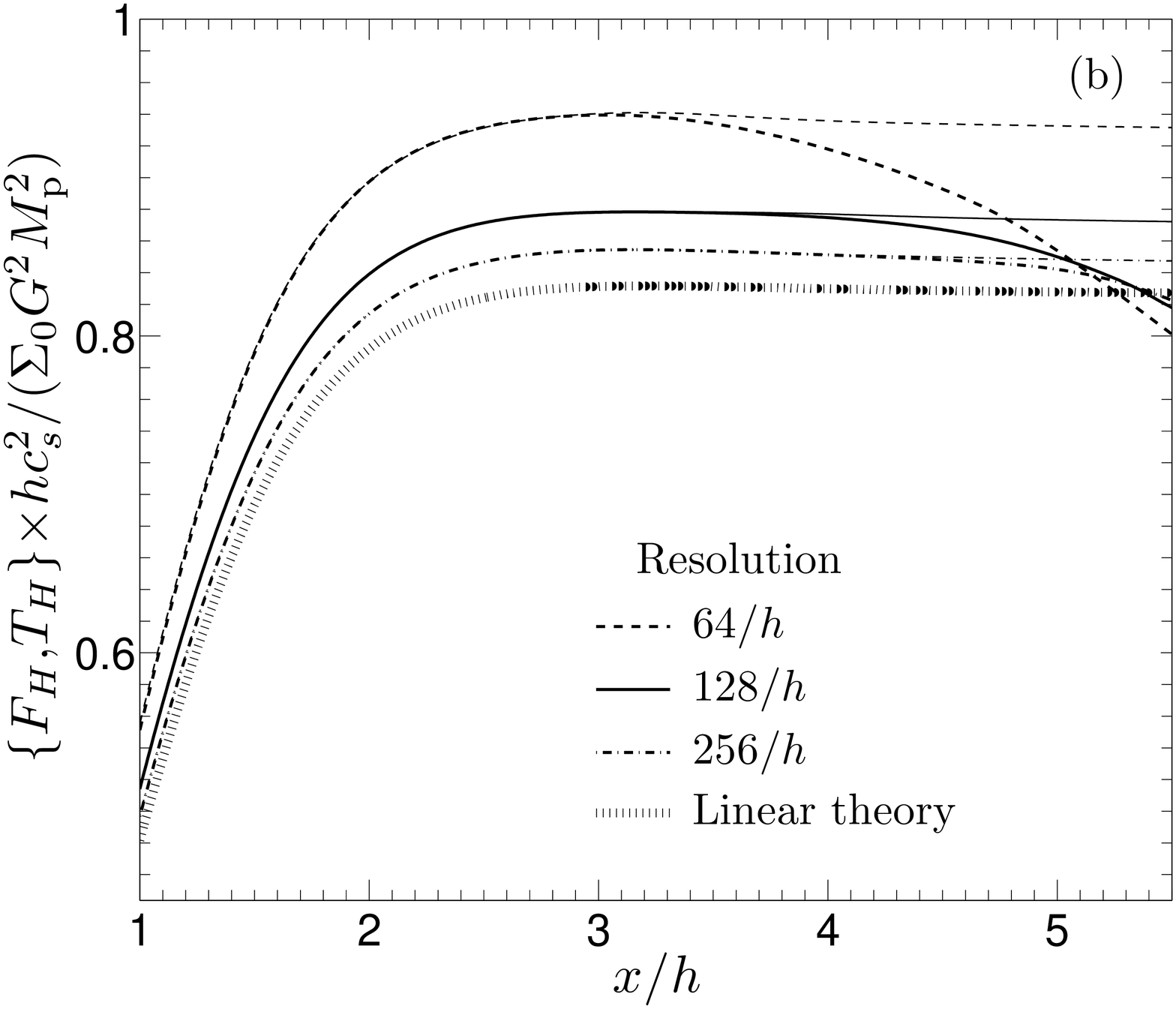} \plotone{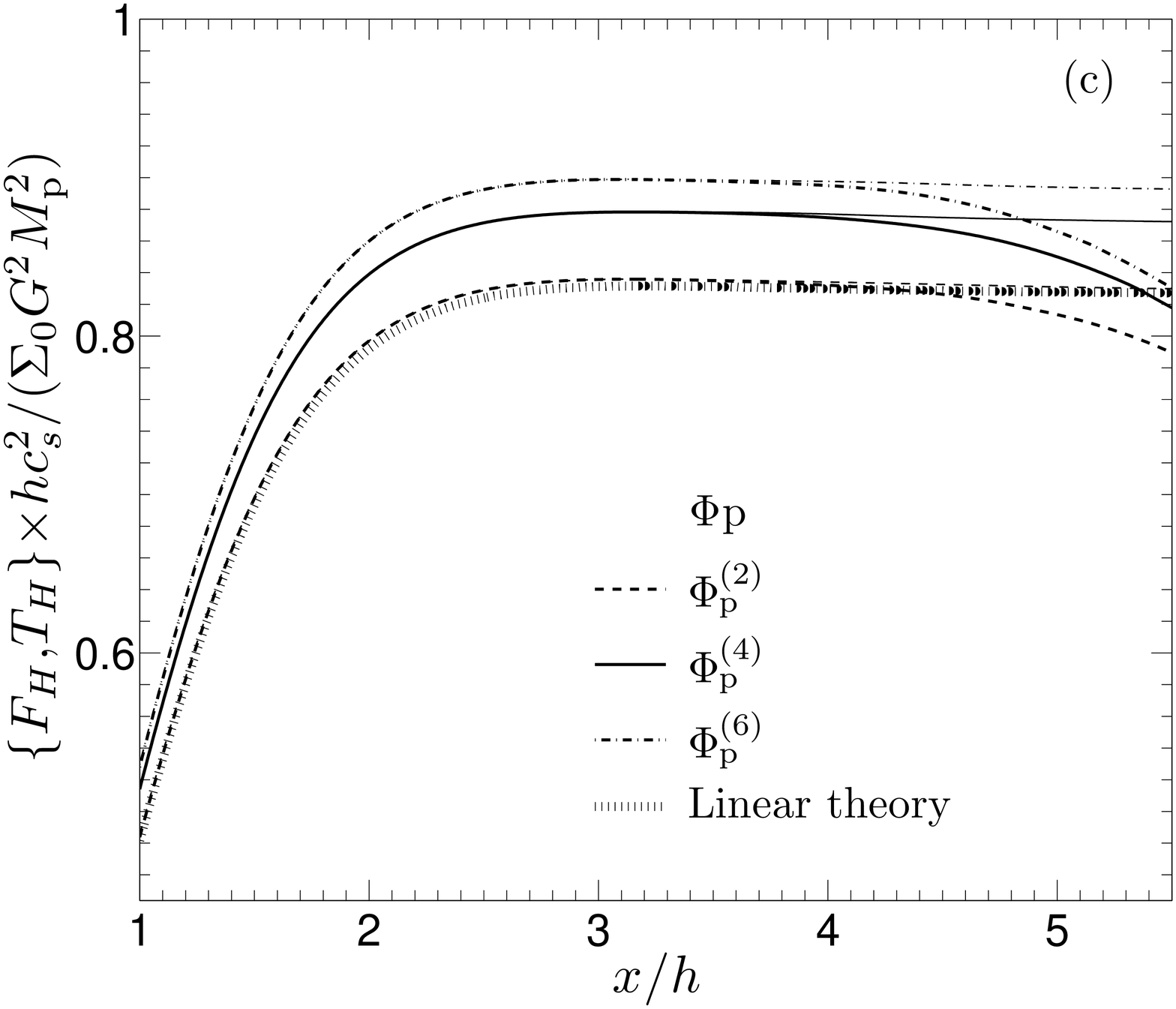} \plotone{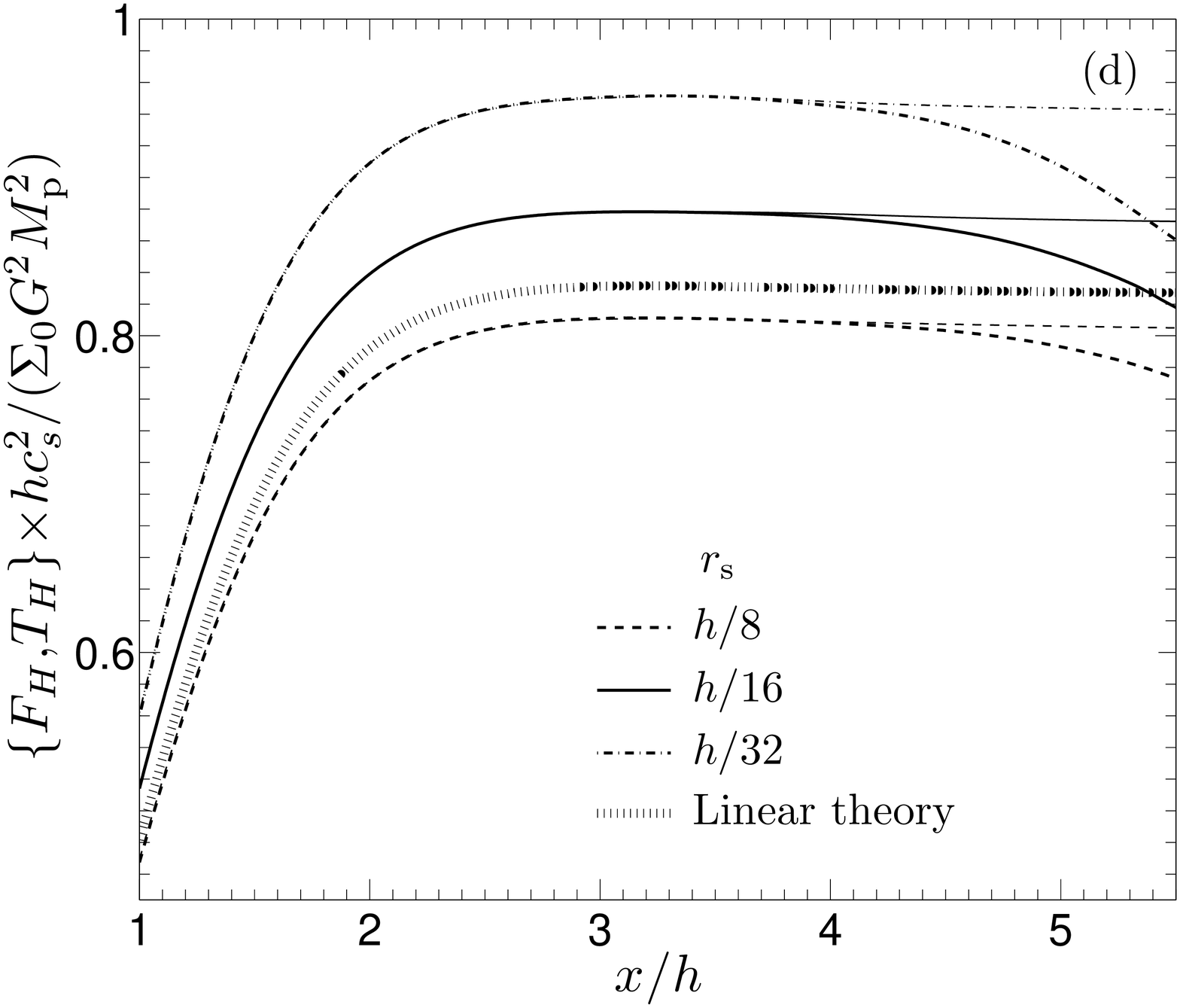}
\end{center}
\figcaption{Same as Figure \ref{fig:density-variety}, but for the spatial behavior of the (properly scaled to remove the dependence on $\mplanet$) integrated torque $T_H(x)$ (the thin line of each line type) and AMF carried by the wave $F_H(x)$ (the thick line of each line type; shifted vertically so that $F_H(0)=0$). The linear theory curve based on the semi-analytical calculation by Rafikov and Petrovich (in preparation) is shown for reference.
\label{fig:amf-variety}}
\end{figure}

In Figure \ref{fig:density-variety}c we show that for a fixed $\rs=h/16$ varying the order of the potential does not have a significant effect on the density wave profile. However, the AMF and torque calculations are more sensitive to the form of the potential, as Figure \ref{fig:amf-variety}c demonstrates. In particular, AMF in calculations with less accurate potential (e.g. $\phip^{(2)}$) is lower than it is in more accurate potential ($5\%$ for $\phip^{(4)}$). We note that the seemingly better agreement with theory in the $\phip^{(2)}$ case is an accidental phenomenon at this set of other numerical parameters. For example, panel (d) shows that using softening length $h/8$ instead of $h/16$ could shift the curves downward by $\sim8\%$, so if switching to $h/8$ in panel (c) then the $\phip^{(6)}$ case would come to perfect agreement with the linear theory (see  \S\ref{sec:sum_res} for additional discussion).

In Figure \ref{fig:density-variety}d we explore the effect of varying $\rs$ for a fixed form of the potential $\phip^{(4)}$. We discover that lowering $\rs$ results in a higher amplitude of the density perturbation. On the other hand, Figure \ref{fig:amf-variety}d shows that a lower $\rs$ leads to a higher numerical AMF and torque, and the difference in the asymptotic value for the two between the $\rs=h/8$ case and the $\rs=h/32$ case is about $16\%$. Again, the better agreement between the theory and the $h/8$ case is accidental. Our calculations presented in \S\ref{sec:result} use $\phip^{(4)}$ potential with $\rs=h/32$.


\subsection{Summary of convergence study}
\label{sec:sum_res}

We conclude that the overall shape of the density wave profile is generally less sensitive to the variations of numerical parameters (unless these values are very different from the ranges explored in this work, see below) than the spatial dependence of the AMF or integrated torque. This emphasizes the importance of these integrated quantities as diagnostics of various subtle effects influencing the wave properties. 

In the course of this numerical exploration we have also discovered that the ``optimal'' value of a particular numerical parameter (e.g. softening length $\rs$) showing closest agreement with the linear theory depends on the values of other numerical parameters (e.g. resolution, type of the potential, etc.) and on $\mplanet$. In other words, there is no universal best choice for each numerical parameter of simulations. In both Figure \ref{fig:density-variety} and \ref{fig:amf-variety}, only the relative position of the curves representing different parameters matters, not their exact locations and the discrepancies between them and the linear theory. On the other hand, as we demonstrate in Paper II, high order of accuracy, high resolution, and highly accurate form of the potential with small softening length are critical to properly resolve the nonlinear stage of the wave evolution.

Hydrodynamic simulations of the disk-planet interaction must often be run for many orbital periods. This is the case {\it e.g.} in studies of planetary migration or gap opening, in which substantial effects appear on timescales of hundreds to thousands of orbital periods (depending on $\mplanet$). This severely restricts the choice of resolution and softening length at which such simulations can run. Typical resolutions in global disk simulations found in the literature \citep{kle99,bry99,nel00,dan02,dan03,li09,yu10,mut10} range from $32/h$ to $1/h$, and $\rs$ is usually taken to be between $0.1h$ to $h$, in combination with a second order of accuracy solver and the second (see Eq. (\ref{eq:phi2})) or even lower order representation of the planetary potential. On the other hand, accurate description of the migration rate and gap opening may only be possible if properties of the density wave excited by the planet are properly captured by the simulation.

\begin{figure}[tb]
\begin{center}
\epsscale{0.45} \plotone{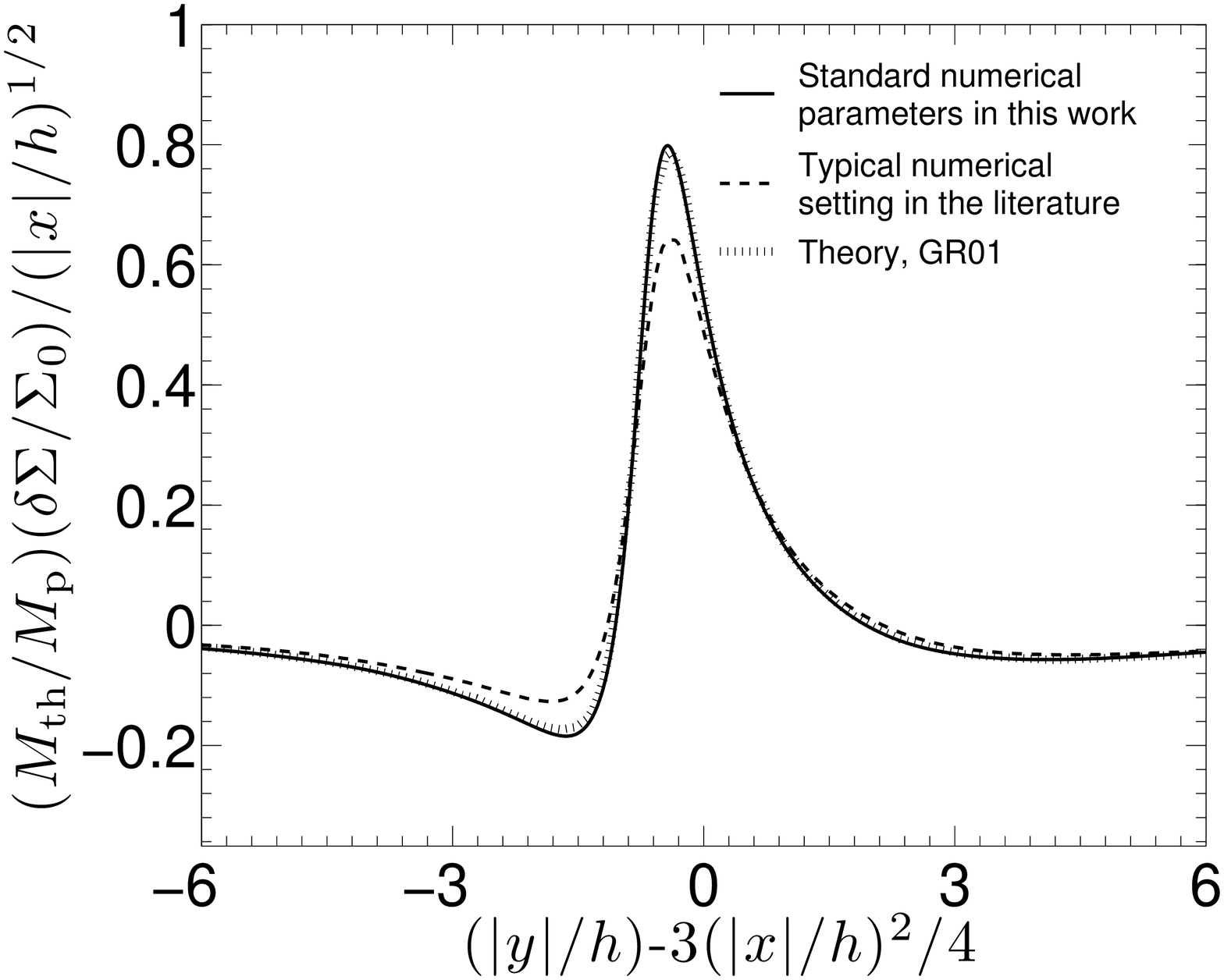} \plotone{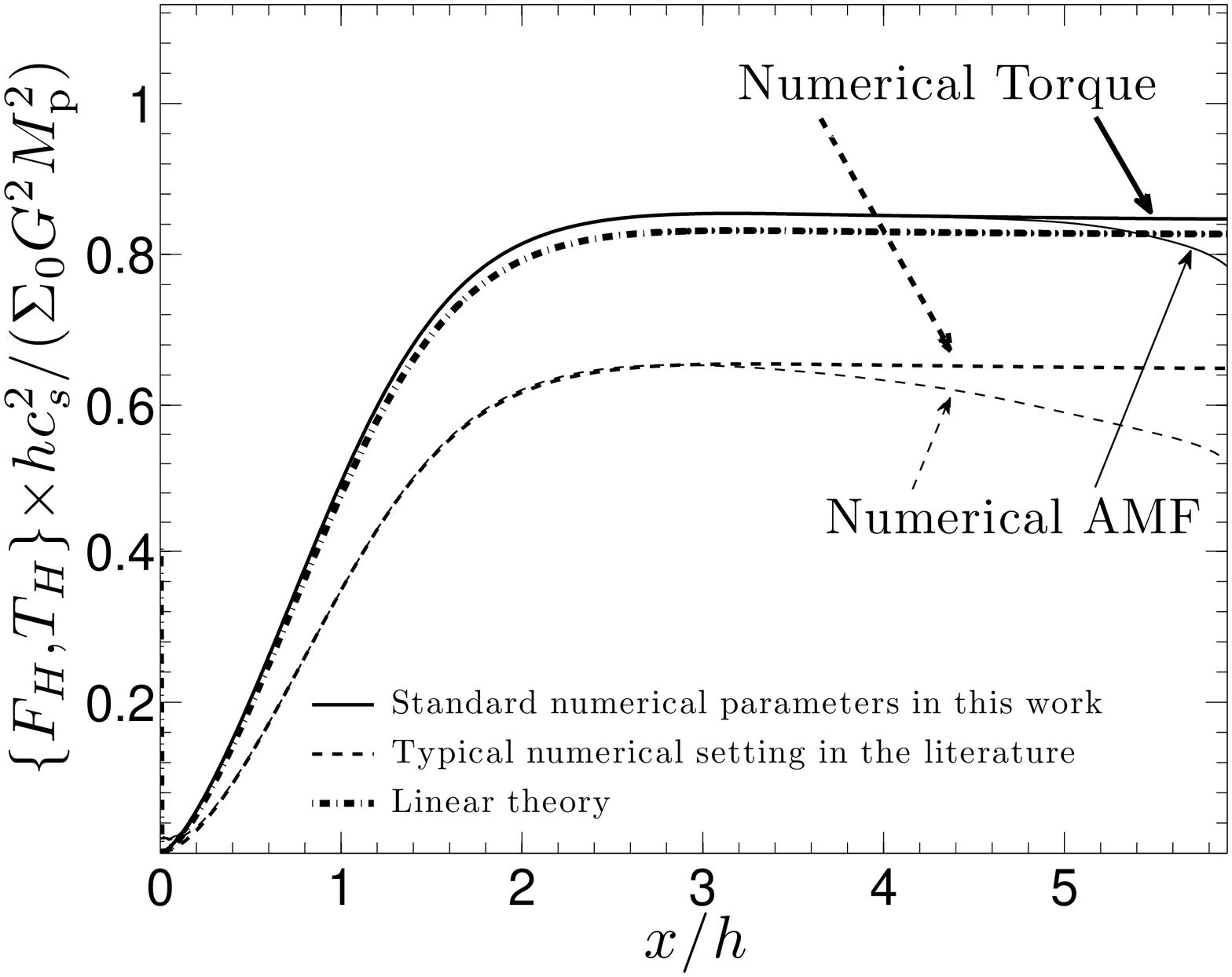}
\end{center}
\figcaption{Comparison between a simulation with our fiducial numerical parameters (solid line, third order of accuracy, resolution $256/h$, $\phip^{(4)}$ potential with $\rs=h/16$), and a simulation with numerical parameters typically adopted in recent planet-disk simulations (dashed line, second order of accuracy, resolution $32/h$, $\phip^{(2)}$ potential with $\rs=h/4$). We use $\mplanet=1.2\times 10^{-2}\Mth$ ($\ls\sim 5h$) for both runs. (a) Azimuthal (scaled) density cut (similar to that in Figure \ref{fig:density-variety}) at $x=1.33h$ showing lower amplitude of the density perturbation in the typical literature run. (b) AMF and integrated torque as a function of $x$ (similar to Figure \ref{fig:amf-variety}) for both runs. The amplitudes of the density perturbation, $F_H(x)$, and $T_H(x)$ are lower, and the AMF dissipation starts earlier in the typical literature case, compared to our fiducial simulation and the linear theory. The linear theory curve is based on the semi-analytical calculation by Rafikov and Petrovich (in preparation).
\label{fig:comparison}}
\end{figure}

To test how reliably the wave structure can be reproduced in global simulations we compare two sets of simulations in our local setting with $\mplanet=1.2\times10^{-2}\Mth$. We run one simulation with the typical numerical parameters for local shearing box simulations found in recent literature (typical global simulations in literature use even smaller resolution and $\rs$): second order of accuracy, resolution $32/h$ and $\phip^{(2)}$ potential with $\rs=h/4$. Another simulation is run with our fiducial parameters --- third order of accuracy, resolution $256/h$, and $\phip^{(4)}$ potential with $\rs=h/16$ --- and their results are compared in Figure \ref{fig:comparison}. 

There are noticeable differences between the two runs. The simulation with typical literature parameters produces smoother density profile, which deviates from analytical solution by about $20\%$, see Figure \ref{fig:comparison}a. At the same time, our fiducial run shows deviations from analytical profile only at the level of $1\%$. 

The difference is even more pronounced when we compare the AMF and torque behavior in Figure \ref{fig:comparison}b. Our fiducial simulation yields asymptotic values of $F_H(x)$ and $T_H(x)$ which are within several per cent of the expected theoretical value (\ref{eq:F_H_tot}). As expected, $F_H(x)$ agrees with $T_H(x)$ all the way until $x\approx 5h$, which is precisely the shock location for the $\mplanet$ used, according to Eq. (\ref{eq:ls}). On the contrary, in the typical literature simulation AMF $F_H(x)$ starts deviating from the integrated torque curve $T_H(x)$ already at $x\approx 2.5h$, significantly in advance of the expected shock position (a factor of 2). This indicates that dissipation and transfer of the angular momentum carried by the wave to the disk fluid start earlier in the typical literature case than they should in reality. The most likely explanation for this behavior is the high level of numerical viscosity in the typical literature run. Moreover, the absolute asymptotic values of $F_H(x)$ and $T_H(x)$ disagree with the theoretical prediction for asymptotic $F_H$ by $\sim23\%$ (comparing with $\sim2\%$ in our fidicual simulation), which is quite significant. We note that the differences between our fiducial and the typical literature simulations persist also in experiments with larger planet masses, $\mplanet\lesssim\Mth$, while the linear regime of wave excitation still holds. 

Underestimate of the wave angular momentum flux in the low-resolution simulations may result in an underestimate of the planetary migration rate in global simulations. In our shearing sheet setup we cannot investigate the effect of resolution on the planetary migration rate: by design one-sided torques exerted by the inner and outer parts of the disk on the planet exactly cancel each other, while the migration speed relies on the small imbalance between them. But the very fact that the one-sided torques in low-resolution case deviate by tens of per cent from the high-resolution case is suggestive that the migration speed should be adversely affected by poor resolution at the same level. Furthermore, the relative imbalance between the one-sided torques can also be a function of resolution potentially exacerbating the problem.

The discrepancy in the AMF should also affect the fidelity of gap opening by the planet in typical literature global simulations. Lower AMF carried by waves means that the planet is less effective at repelling gas away from its semi-major axis, which would require higher $\mplanet$ to open a gap. In addition, the gap opening process is sensitive to the {\it spatial distribution} of the AMF dissipation \citep{R02b}. As a result, spurious dissipation of wave AMF prior to shock formation in low-resolution simulations and the accelerated AMF decay after the shock may introduce artificial effects in the gap opening picture. To summarize, any numerical studies of processes in which density wave production and dissipation plays important role need to use high order of accuracy, high resolution, and accurate representation of the planetary potential with small softening length.

\section{A numerical issue in planet-disk simulations}\label{sec:fargo}

In this section we describe a commonly ignored numerical problem that we discover in disk-planet interaction simulations. Namely, we find that if the time-step $dt$ of the simulation is too large, the code cannot resolve the motion of the fluid in regions where the fluid experiences large gravitational acceleration. This issue applies to simulations in general, but it is especially problematic when the orbital advection algorithms is implemented (the FARGO algorithm, citealt{mas00}).

\citet{mas00} introduced a modification of the standard transport algorithm (Fast Advection in Rotating Gaseous Objects, FARGO) for simulations of differentially rotating systems, which significantly speeds up the calculations. By getting rid of the average azimuthal velocity when applying the Courant condition, this technique results in a much larger time step $dt$, which is limited by the small perturbed velocity, than the usual procedure, where $dt$ is limited by the full fluid velocity dominated by the differential rotation. FARGO has been implemented in Athena by \citet{sto10}.

In a shearing box without planets, the dynamical time scale is $\Omega^{-1}$ and is uniform throughout the box.  However, when the planet is introduced, the dynamical timescale is spatially varying, and could be characterized by the free-fall time, which is the time that it takes a fluid element to fall on the planet assuming a constant acceleration at the grid point. For gravitational potential in the form of Eq. (\ref{eq:phi4}), the free-fall time is:
\begin{equation}
t_{ff}=\sqrt[]{\frac{2(\rho^2+\rs^2)^{5/2}}{{\rm G}\mplanet(\rho^2+2.5\rs^2)}}
\label{eq:tff}
\end{equation}
Note that $t_{ff}$ depends on a specific form of the gravitational potential, and a more smoothed potential results in a larger $t_{ff}$. The smallest free-fall time in the entire domain, which is the limiting timescale for the simulation, occurs at the grid points that are adjacent to the planet, which have the smallest $\rho$ ($1/\sqrt[]{2}$ of the cell size $r_c$ in Athena; the separation where the smallest $t_{ff}$ occurs also depends on the form of potential). In our simulations, we always keep $\rs^2\gg r_c^2$ (usually $\rs=8r_c$), so the smallest free-fall time is:
\begin{equation}
t_{ff,s}\approx\sqrt[]{\frac{0.8\rs^3}{{\rm G}\mplanet}}
\label{eq:tffs}
\end{equation}
On the other hand, according to the Courant condition when varying the resolution the time step $dt\propto r_c$. So the ratio $t_{ff,s}/dt$ scales with numerical parameters and $\mplanet$ as
\begin{equation}
\frac{t_{ff,s}}{dt}\propto\frac{\rs^{1.5}}{\sqrt[]{\mplanet} r_c}
\label{eq:tffdt}
\end{equation}
To properly resolve fluid dynamics in the vicinity of the planet, the ratio $t_{ff,s}/dt$ should be kept above a certain threshold.

Our calculations without orbital advection and with $dt$ determined by the Courant condition usually have $t_{ff,s}/dt\sim150$. However when we turn on the orbital advection algorithm, $dt$ typically increases by a factor of $\sim10$, and the ratio $t_{ff,s}/dt$ decreases by the same factor. We find that when the time-step is {\it too large} to properly resolve the fluid motion around the planet, the numerical results will be {\it incorrect}, as described below.

\begin{figure}[tb]
\begin{center}
\epsscale{0.5} \plotone{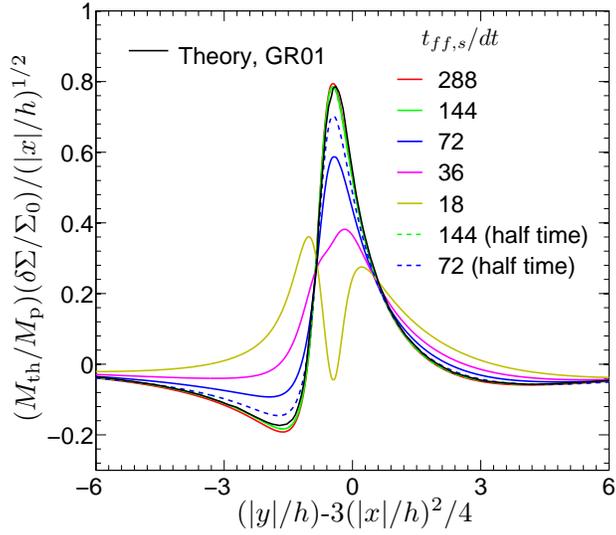}
\end{center}
\figcaption{Azimuthal density profiles $\delta\Sigma$ (scaled by the planetary mass and normalized by $(x/h)^{1/2}$) at $x=1.33h$, for simulations using orbital advection algorithm (FARGO) with different $t_{ff,s}/dt$ ratio (we manually set $dt$ in different cases to be a fraction of the value of $dt$ set by the Courant condition in FARGO). Dashed lines show the density profiles at half simulation time for two representative runs to illustrate whether temporal convergence has been achieved. Theoretical prediction (GR01) is also plotted to guide the eye. Simulations are done with resolution $64/h$, $\phip^{(4)}$ potential with $\rs=h/8$, and $\mplanet=3.2\times10^{-2}\Mth$.
\label{fig:fargo}}
\end{figure}

Figure~\ref{fig:fargo} shows the azimuthal density profiles $\delta\Sigma$ at $x=1.33h$ for a set of simulations using identical numerical parameters and orbital advection algorithm but with different $t_{ff,s}/dt$ ratio, which we achieve by manually setting $dt$ to be a fraction of the $dt$ set by the Courant condition in FARGO. The two cases with highest $t_{ff,s}/dt$ yield the density profile in agreement with the theoretical prediction, also demonstrating the convergence of the result at high $t_{ff,s}/dt$. However, the density profiles in simulations with lower $t_{ff,s}/dt$ clearly deviate from theory, with smaller $t_{ff,s}/dt$ leading to larger discrepancy. In the run with the smallest used $t_{ff,s}/dt=18$ which corresponds to $dt$ set by the Courant condition in FARGO (representing maximum FARGO speed up), the density peak even becomes a density trough. By experimenting with Athena, we generally find that our simulations of the disk-planet interaction produce converged results agreeing with theory only when
\begin{equation}
t_{ff,s}/dt\gtrsim100,
\label{eq:tffdtcondition}
\end{equation}
while this critical number may depend on the numerical method used in a particular code or some other numerical conditions. We note that in principle all disk-planet simulations have to satisfy condition (\ref{eq:tffdtcondition}) to ensure correct results, but in experiments we find simulations with the orbital advection algorithm tend to violate this condition much more easily than simulation without the orbital advection algorithm.

Apart from incorrectly reproducing density perturbation, disk-planet simulations violating the $t_{ff,s}/dt$ constraint exhibit other serious problems. In particular, simulations with small $t_{ff,s}/dt$ do not achieve a steady state. We illustrate this in Figure~\ref{fig:fargo} by showing the density profile at half simulation time for two cases. While the $t_{ff,s}/dt=144$ case reaches a steady state and develops time independent density profile at half time, the $t_{ff,s}/dt=72$ case does not, and its density perturbation decreases with time.

In simulations with $t_{ff,s}/dt$ lower then the critical value, the fluid element approaching the planet at small $|x|$, instead of moving on a horseshoe orbit, gets trapped by the planetary potential, leading to the formation of a fluid loop around the planet. As simulation progresses the rotational velocity of the fluid loop becomes higher and higher, and eventually the centrifugal force evacuates the region around the planet. This effect, which is a purely numerical artifact of the too small $t_{ff,s}/dt$, is visually very similar to (and can easily be confused with) the gap opening phenomenon, which is a real physical effect. Simulations exploring the gap opening process with large planet mass and low resolution are likely to have small $t_{ff,s}/dt$ violating condition (\ref{eq:tffdtcondition}, see Eq. \ref{eq:tffdt}), which may have detrimental effects on the results. We note that using Athena with orbital advection algorithm, numerical simulations with $\mplanet\sim\Mth$ and a large $\rs=h/4$ must have an effective resolution at least $64/h$ in the vicinity of the planet to avoid this problem (a more smoothed potential may weaken this condition, through). However, again we emphasize that the critical value of $t_{ff,s}/dt$ in simulations with the same numerical parameters but using different code and solver may be different from ours.

The $t_{ff,s}/dt$ threshold severely limits the ability of FARGO to speed up calculations of disk-satellite interaction. For most of our simulations, $t_{ff,s}/dt$ is already rather close to the threshold value without orbital advection algorithm, so there is not much room left to increase $dt$, which is what the orbital advection algorithm aims to achieve (all our simulations presented outside this section are done without orbital advection technique). However, in many other studies of differentially rotating systems, such as the investigation of magnetorotational instability (MRI) in accretion disks, the point mass objects are absent and the characteristic dynamical time scale is always long enough for the orbital advection algorithm to be a extremely useful tool for speeding up the simulations. At last, we note that both $dt$ and $t_{ff,s}$ may vary during the simulation, and the constraint (\ref{eq:tffdtcondition}) on their ratio should be satisfied in the entire domain throughout the simulation time.


\section{Summary}
\label{sec:summary}


We have conducted a series of hydrodynamical simulations to study the details of the gravitational interaction between low mass planets and a protoplanetary disk, and to test predictions of the linear theory of density wave evolution. Our simulations assume local shearing sheet geometry and are carried out in 2D at very high resolution to reduce the effect of numerical viscosity. We focus on both the excitation of the density waves and their propagation away from the planet in the linear regime. To mitigate the effects of nonlinearity we consider very small planetary masses, starting at $0.4$ M$_\oplus$ and going down to 3 Lunar masses at 1 AU. 

We extract the spatial distribution of the density perturbation induced by the planet from our simulations and compare it with the predictions of linear theory. We find good agreement between the two, at the level of several per cent when high resolution (typically $256/h$) is employed. We also investigate the spatial dependence of the angular momentum flux carried by the wave and the distribution of torque induced by the planet on the disk, again finding good agreement with theoretical predictions. In particular, we are able to reproduce the theoretical behavior of the torque cutoff in Fourier space. 

We also find various manifestations of nonlinear effects in our simulations even while the waves are formally linear. These include (1) the noticeable steepening of the density profile at larger values of $\mplanet$ far from the planet, (2) the slight variation of the asymptotic value of the AMF with planetary mass in addition to the expected $F_H\propto \mplanet^2$ dependence, causing the discrepancy with linear theory prediction at the level of several per cent, and (3) the growth of power in high azimuthal wavenumber harmonics of the AMF in Fourier space.

By carefully studying the spatial distribution of the torque density $dT_H(x)/dx$ in our simulations we discover an interesting ``negative torque'' phenomenon: $dT_H(x)/dx$ changes sign at some radial separation from the planet (at $|x_-|\approx 3.2h$ in our simulations), which contradicts the analytical results of GT80. This effect can however be understood in the framework of the linear theory as shown in Rafikov \& Petrovich (in preparation). This feature of the torque distribution has only weak effect on the total accumulated torque in our simulations.

We also carried out a detailed investigation of the effect of different numerical settings on our results in linear regime. We explored the influence on wave properties of (1) different Riemann solvers with different accuracy, (2) spatial resolution, (3) different forms of the softened planetary potential, and (4) softening length of the planetary potential. We find the spatial distribution of the AMF and torque to be a more sensitive probe of the effects of various numerical parameters on the wave evolution than the distortions of the density wave profile. Based on this study we conclude that a very high resolution (ensuring low numerical viscosity), high order of accuracy, and an accurate prescription of planetary potential with small softening length are critical for accurately reproducing the key features of the wave evolution in linear regime.

We demonstrate that low order of accuracy, low spatial resolution, and inaccurate potential with large softening length often employed in global disk-planet interaction studies can severely affect the fidelity of their results, especially in applications to planet migration and gap opening. Specifically, we conduct a test run with typical numerical parameters from recent literature. Comparing with the analytical theory, the test run produces a lower amplitude of the density wave (by $\sim20\%$), a lower final accumulated torque (by $\sim23\%$), and a largely advanced starting point of wave dissipation (a factor of $\sim2$). 

Most of our own calculations were carried out with third order of accuracy, resolution of $256/h$, softening length $\rs=h/32$ and a potential that rapidly converges to Newtonian (as $(\rs/\rho)^4$) at large separation $\rho$ from the planet. This set of numerical parameters allows us to obtain excellent agreement with linear theory in Athena, but is not meant to be universal for other codes. However, the way in which we make the comparison with the theory and the agreement we achieve may serve as a standard framework for future code tests.

We also discover a numerical problem which is largely ignored in previous simulations of disk-planet interactions. To follow the fluid motion correctly, the time-step in a simulation has to be smaller than the local dynamical timescale by a certain factor ($\sim100$ in our case using Athena) in the entire domain, including the region where the fluid experiences the largest gravitational acceleration ({\it e.g.} the vicinity of the planet). In the context of numerical disk-planet studies, violation of this condition leads to inaccurate calculation of wave properties, lack of convergence on long time scales, and spurious repulsion of gas by planet similar to gap opening. This timescale issue applies to disk-planet simulations in general, but it particularly affects the ones which use the orbital advection algorithm (FARGO), in which cases a significant increase of the time step is usually allowed to speed up calculations.

In Paper II, we will continue our investigation of the disk-planet interaction by looking at the details of the density wave evolution in the nonlinear phase.

\section*{Acknowledgments}

We are grateful to Jeremy Goodman, Takayuki Muto, Zhaohuan Zhu, and an anonymous referee for useful discussions and comments. The financial support of this work is provided by NSF grant AST-0908269 and Sloan Fellowship awarded to RRR.


\clearpage

\end{document}